\documentclass[aps,prl,twocolumn,superscriptaddress,floatfix,citeautoscript,preprintnumbers,footinbib]{revtex4-1}
\usepackage[dvipsnames]{xcolor}
\usepackage{amsmath,amssymb}
\usepackage{commath}
\usepackage{todonotes}
\usepackage[colorlinks = true,
            linkcolor = blue,
            urlcolor  = blue,
            citecolor = blue,
            anchorcolor = blue]{hyperref}

\usepackage{mathtools}            
\usepackage{cleveref}
\usepackage{dsfont}
\usepackage{graphicx}
\begin{document}
\crefname{equation}{Eq.}{Eqs.}
\crefname{figure}{Fig.}{Fig.}
\crefname{appendix}{Appendix}{Appendix}

\newcommand{\dan}[1]{\textcolor{BrickRed}{DM: #1}}
\newcommand{\ket}[1]{|#1\rangle}

\newtheorem{proposition}{Proposition}

\title{Sequential generation of projected entangled-pair states}

\author{Zhi-Yuan Wei}

\thanks{These two authors contributed equally.}
\author{Daniel Malz}

\thanks{These two authors contributed equally.}
\author{J. Ignacio Cirac}

\affiliation{Max-Planck-Institut f{\"u}r Quantenoptik, Hans-Kopfermann-Stra{\ss}e 1, D-85748 Garching, Germany}

\affiliation{Munich Center for Quantum Science and Technology (MCQST), Schellingstr. 4, D-80799 M{\"u}nchen, Germany}

\date{\today}

\begin{abstract}
We introduce plaquette projected entangled-pair states, a class of states in a lattice that can be generated by applying sequential unitaries acting on plaquettes of overlapping regions. They satisfy area-law entanglement, possess long-range correlations, and naturally generalize other relevant classes of tensor network states. We identify a subclass that can be more efficiently prepared in a radial fashion and that contains the family of isometric tensor network states [M. P. Zaletel and F. Pollmann, Phys. Rev. Lett. 124, 037201 (2020)]. We also show how this subclass can be efficiently prepared using an array of photon sources.
\end{abstract}
          
\maketitle
Tensor network states play a fundamental role both in quantum information processing and many-body physics, as they are natural representations of states with area-law entanglement~\cite{verstraete2008matrix,Orus2019,Cirac2020}. In one dimension, Matrix-Product States (MPS)~\cite{fannes1992finitely,Perez-Garcia2006,schollwock2011density} efficiently approximate the ground state of gapped~\cite{hastings2007area} and critical Hamiltonians~\cite{Verstraete2006c}.
Their higher-dimensional generalizations, Projected Entangled-Pair States (PEPS)~\cite{verstraete2004renormalization}, also play an important role in many-body physics. Apart from providing efficient approximations in different scenarios, they embrace many paradigmatic states of condensed matter physics, including topological states like the toric code~\cite{Verstraete2006,Schuch2010} and string-net states~\cite{levin2005string,gu2009tensor,Buerschaper2009}, or resonating valence bound states~\cite{Verstraete2006}. 
They also contain elements that are relevant in the context of quantum metrology~\cite{Degen2017}, like the W~\cite{dur2000three} or GHZ states~\cite{greenberger1989going}, or in quantum computing, like the cluster~\cite{Briegel2001}, graph~\cite{PhysRevA.69.062311,Russo2019,Azuma2015} and hypergraph state~\cite{kruszynska2009local,Rossi2013}. Thus, the efficient preparation of such states would have an important impact on the study of many-body systems and quantum information.

One can generate MPS by sequentially applying local unitaries~\cite{Schon2005,Schon2007}, which provides a way to deterministically prepare entangled states on quantum computers~\cite{Smith2019,lin2021real} or in photonic systems~\cite{gheri1998entanglement,saavedra2000controlled,Schon2007,lindner2009, schwartz2016, tiurev2020a, Eichler2015, schwartz2016,Besse2020, zyryd,Wein2021}, with a generation time (circuit depth) that scales linearly with the system size $n$ (number of qudits) as $O(n)$. Moreover, sequential MPS generation is an essential component in numerous theoretical frameworks~\cite{Lamata2008,Cramer2010a,Saberi2011,delgado2007sequential,Osborne2010,Huggins2019,Foss-Feig2020a,Barratt2021,Ran2020a,lin2021real}.

Efficient generation of PEPS is, however, much more difficult. Even in two dimensions, it is believed that most states will require a preparation time that increases exponentially with the system size~\cite{Verstraete2006,Schuch2007}.
Nevertheless, most of the paradigmatic examples mentioned above can be efficiently prepared also in higher dimensions, and experimental efforts have already started~\cite{satzinger2021,yj_circ}.
This calls for efforts to identify, classify, and extend subclasses of PEPS that allow for efficient preparation, ideally together with an explicit algorithm to do so.
In this vein, there are two subclasses of PEPS in two dimensions that stand out: (i) sequentially generated states (SGS)~\cite{Banuls2008}, and (ii) PEPS generated by photon feedback (F-PEPS)~\cite{Pichler2017}. Interestingly, both of these classes can be obtained from a product state by a sequential quantum circuit.

In this paper, we introduce \textit{plaquette} PEPS (P-PEPS), which are defined by sequentially applying unitaries to plaquettes of qudits initially in a product state.
P-PEPS can straightforwardly be expressed as PEPS and naturally encompass SGS and F-PEPS. 
We focus on a particular radial plaquette ordering, which leads to a subclass we call \textit{radial plaquette} PEPS (RP-PEPS).
This class allows certain local observables to be computed efficiently and has SGS and isometric tensor network states (isoTNS)~\cite{Zaletel2020} as proper subclasses.
Our construction thus provides a quantum circuit to prepare isoTNS, which is a class that has been shown to include graph states and hypergraph states of local connectivities, and all string-net states~\cite{Soejima2020}.
While for a $q$-dimensional lattice of $N = n_1 \times ... \times n_q$ sites, in the worst case, P-PEPS require a circuit depth scaling with the total number of sites,
RP-PEPS can be prepared particularly efficiently, with the circuit depth $T_{\rm rp}$ scales as the side length of the lattice
\begin{equation} \label{radial_speed}
{T_{{\rm{rp}}}} = O( {\mathop {\max }\limits_i {n_i}} ).
\end{equation}

We also show that an array of coupled quantum sources each comprising an ancilla--emitter pair can naturally produce RP-PEPS of flying qub(d)its with the same efficient scaling, and prepare F-PEPS with a circuit depth $O(N)$. This includes a wide variety of high-dimensional states that have been proposed for sequential photon generation~\cite{Economou2010,Gimeno-Segovia2019,Russo2019,Bekenstein,Pichler2017,Dhand2018,Xu2018,Wan2020,Zhan2020,Shi2021,Bartolucci2021,Bombin2021,zyryd}. Overall, P-PEPS (RP-PEPS) and their generation protocols apply to photonic systems, and to platforms with matter qub(d)its like superconducting circuits~\cite{blais2021circuit}, trapped ions~\cite{Bruzewicz2019}, or Rydberg atoms~\cite{saffman2016quantum}, where local interactions can be engineered with high precision.

\emph{Plaquette PEPS.---}For concreteness, we restrict our attention to a two-dimensional lattice of qudits of size $N = n \times m$, and the high-dimensional generalization will be discussed in SI~\footnote{See Supplemental Material for more details, which includes Refs.[74,75]}.

\begin{figure}[t]
	\centering
	\includegraphics[width=0.48\textwidth]{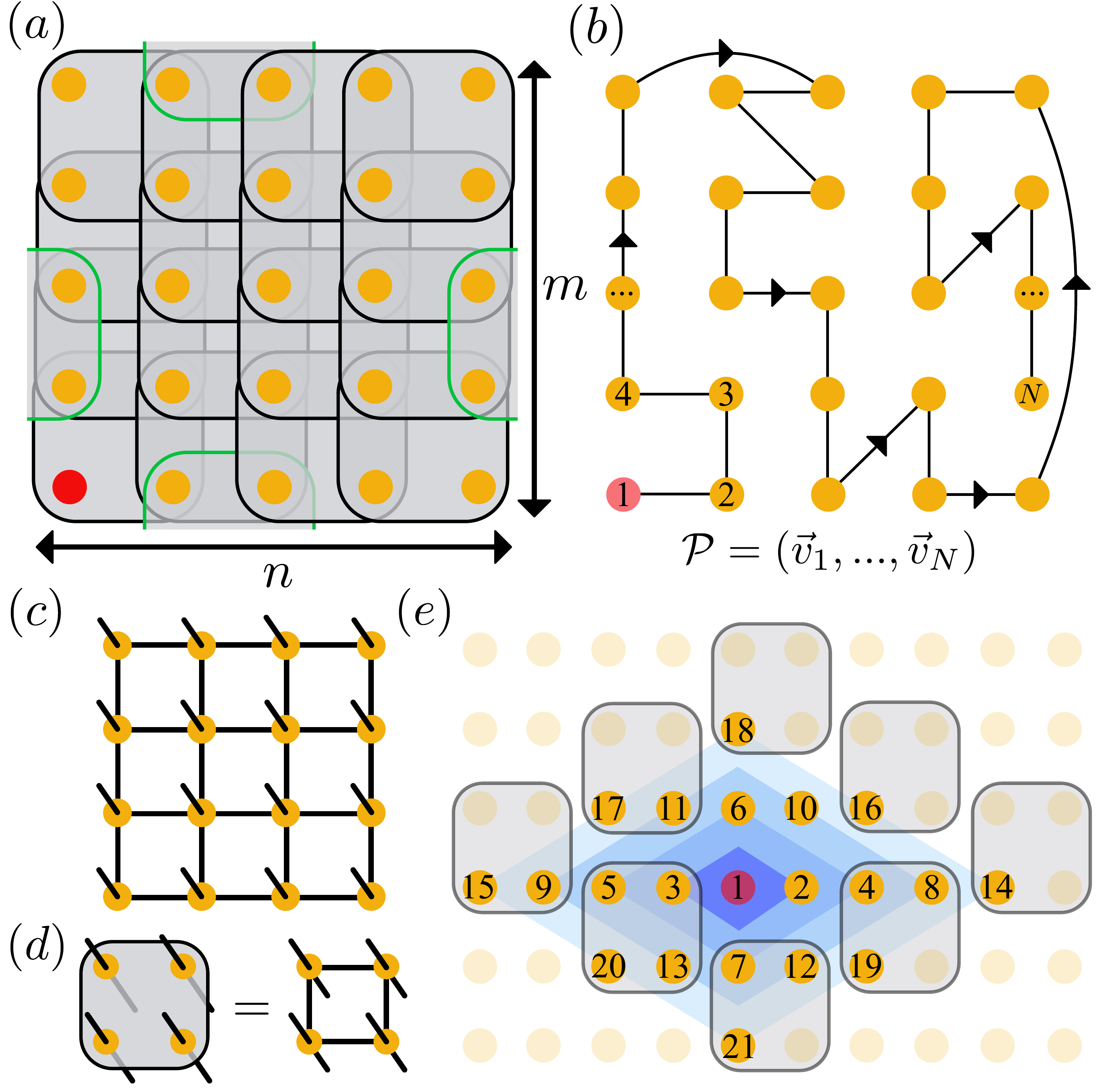}
        \caption{(a) Plaquette PEPS (P-PEPS) are prepared by sequentially applying plaquette unitaries $\{\hat U_{\vec v_{\mu}}\}$ (denoted by the gray squares) of size ${L_p} \times {L_p}$ ($L_p=2$ here) to a product state. The source point $\vec v_1$ is marked by a red dot.
        Periodic and open boundary conditions are distinguished by whether or not the sequence of unitaries contains that act across the boundary (green squares).
(b) P-PEPS is determined by the unitaries and their ordering $\cal P$. Here we show an example of $\cal P$ using a directional string and the numbers from 1 to $N$. (c) PEPS are states defined through connected networks of tensors with virtual indices (connected lines) of bond dimension $D$, and physical indices of dimension $d$ (sticking out).
(d) The plaquette unitaries can be decomposed into PEPO, which lead to a PEPS representation of P-PEPS.
(e) Preparation of a \textit{radial} plaquette PEPS with $L_p=2$. Here starting from the source point (the red dot), the ordering $\cal P$ is denoted by the numbers. Such ordering allows one to apply each layer (denoted by the shades of different colors) of unitaries in parallel. The gates in the 5-th layer are denoted by gray squares.}
        \label{P-PEPS_def_fig}
\end{figure}

We define P-PEPS with periodic boundary conditions (PBC) as the states generated from the product state $\ket0^{\otimes N}$ through sequential application of unitaries to plaquettes of size ${L_p} \times {L_p}$ (${L_p} \ll m,n$) [c.f.~\cref{P-PEPS_def_fig}(a)]
\begin{equation} \label{P-PEPS}
\left| {{\psi _{{\rm{p}}}}} \right\rangle  = \prod\limits_{\mu = 1}^N {{{\hat U}_{\vec v_\mu}}} {\left| 0 \right\rangle ^{ \otimes N}},
\end{equation}
where $\vec v_\mu=(i_\mu,j_\mu)$ and the unitary $\hat U_{\vec v_{\mu}}$ acts on qudits in the square spanning from $(i_{\mu},j_{\mu})$ to $(i_{\mu}+L_p-1,j_{\mu}+L_p-1)$, and we identify the rows $i \pm m \equiv i$ and columns $j \pm n \equiv j$. Here the choice of plaquette shape reflects the locality of the unitaries.
The ordering of the unitaries ${\cal P} = \left( {{{\vec v}_1},{{\vec v}_2},...,{{\vec v}_N}} \right)$ fulfils the conditions $\vec v_\mu \neq  \vec v_\nu$ for $\mu \neq \nu$. We show an example of $\cal P$ in \cref{P-PEPS_def_fig}(b), and call the position of the first unitary $\vec v_1$ the \textit{source point}. To define the state with open boundary conditions (OBC), we simply omit gates that act across boundaries.

P-PEPS is a subclass of PEPS, which are states defined through a network of tensors with one tensor per lattice site [\cref{P-PEPS_def_fig}(c)],
whose virtual indices are contracted with their neighbors.
In two dimensions,
\begin{equation} \label{peps_def}
    |\Psi _{\rm PEPS} \rangle  = \sum\limits_{\{ k\}  = 0}^{d - 1} {{{\cal F}_{{\rm{2D}}}}(\{ {B_{[ {i,j} ]}^k}_{lurb}\} )|\{ k\} \rangle } ,
\end{equation}
where ${B_{[ {i,j} ]}^k}_{lurb}$ is a rank-5 tensor on the site $(i,j)$ that has one physical index $k$ of dimension $d$ and four virtual indices $l,u,r,b$ of \textit{bond dimension} $D$.
The symbol $\cal F_{\rm 2D}$ denotes the contraction of all virtual indices.
To obtain the PEPS representation of P-PEPS, we decompose the plaquette unitaries into projected entangled-pair operators (PEPO)~\cite{Cirac2020} [c.f.~\cref{P-PEPS_def_fig}(d)]. This allows one to write the whole sequential circuit as a PEPO, which, applied to a product state, yields a PEPS with bond dimension $D\le O(d^{L_p^4})$~\cite{Note1}.

We are particularly interested in cases where each unitary overlaps with at least one of the earlier ones, such that they create correlations. The state shown in \cref{P-PEPS_def_fig}(a) is such an example. Sequential circuits with overlapping unitaries efficiently establish correlations between arbitrary locations of the lattice with $O(N)$ unitaries~\cite{Note1}.
This should be contrasted with brickwall circuits~\cite{Gopalakrishnan2019} that take $O(N\cdot \mathop {\max }\limits_i {n_i})$ unitaries to do so~\cite{Note1}.
This implies that P-PEPS offers a more efficient parametrization of states with correlations across the entire system.
Moreover, while P-PEPS have area-law entanglement, brickwall circuits that create long-range correlations will instead lead to states with volume-law entanglement~\cite{Nahum2017,Nahum2018,VonKeyserlingk2018}.

\emph{Radial plaquette PEPS (RP-PEPS).---} Naively, it takes a circuit depth $O(N)$ to create a P-PEPS [c.f.~ \cref{P-PEPS}]. However, some orderings $\cal P$ allow unitaries to be applied in parallel. A simple example is to arrange the unitaries as a brickwall circuit of depth $O(L_p^2)$.
Here we define a subclass of P-PEPS, RP-PEPS, where starting from the source point, the positions $\{ \vec v_{\mu} \}$ of the unitaries are ordered such that they can be grouped to multiple layers of commuting unitaries, and each layer act on the boundary of the existing gate-acted region. An example with $L_p=2$ is illustrated in \cref{P-PEPS_def_fig}(e), where the gates are grouped as $\left[ {\left( 1 \right),\left( {2,3} \right),\left( {4 \textrm{--} 7} \right),\left( {8 \textrm{--} 13} \right),\left( {14 \textrm{--} 21} \right),...} \right]$ (denoted by shades of different colors). To resolve ambiguities in the plaquette order, we choose \textit{preferred directions} in which the position of the plaquette moves. In \cref{P-PEPS_def_fig}(e) we choose `horizontal first, and positive direction first'. The circuit depth of preparing RP-PEPS is asymptotically ${T_{{\rm{rp}}}} \approx n + {L_p}m$, 
 following the scaling in \cref{radial_speed}. Moreover, RP-PEPS allow efficient computation of expectation values of local observables that are geometrically close to the source point or the line that passes through the source point along the preferred direction, and this generically implies that correlation functions in these regions decay exponentially. This is reminiscent of isoTNS~\cite{Zaletel2020}.

The above definitions straightforwardly generalize to higher dimensional lattices, where plaquettes become high-dimensional cubes~\cite{Note1}. While the general circuit depth for P-PEPS again scales with $N$, RP-PEPS obeys \cref{radial_speed}.

\emph{Relation to other families of PEPS.---}By definition, P-PEPS can be efficiently prepared, have a PEPS description, and host long-range correlations. 
Now we show that P-PEPS naturally encompass other families of PEPS that are prepared sequentially (SGS and F-PEPS), as well as isoTNS (we follow the definition in Ref.~\cite{Zaletel2020}, and see Ref.~\cite{Haghshenas2019} for a different definition).

\begin{figure}[t]
	\centering
	\includegraphics[width=\linewidth]{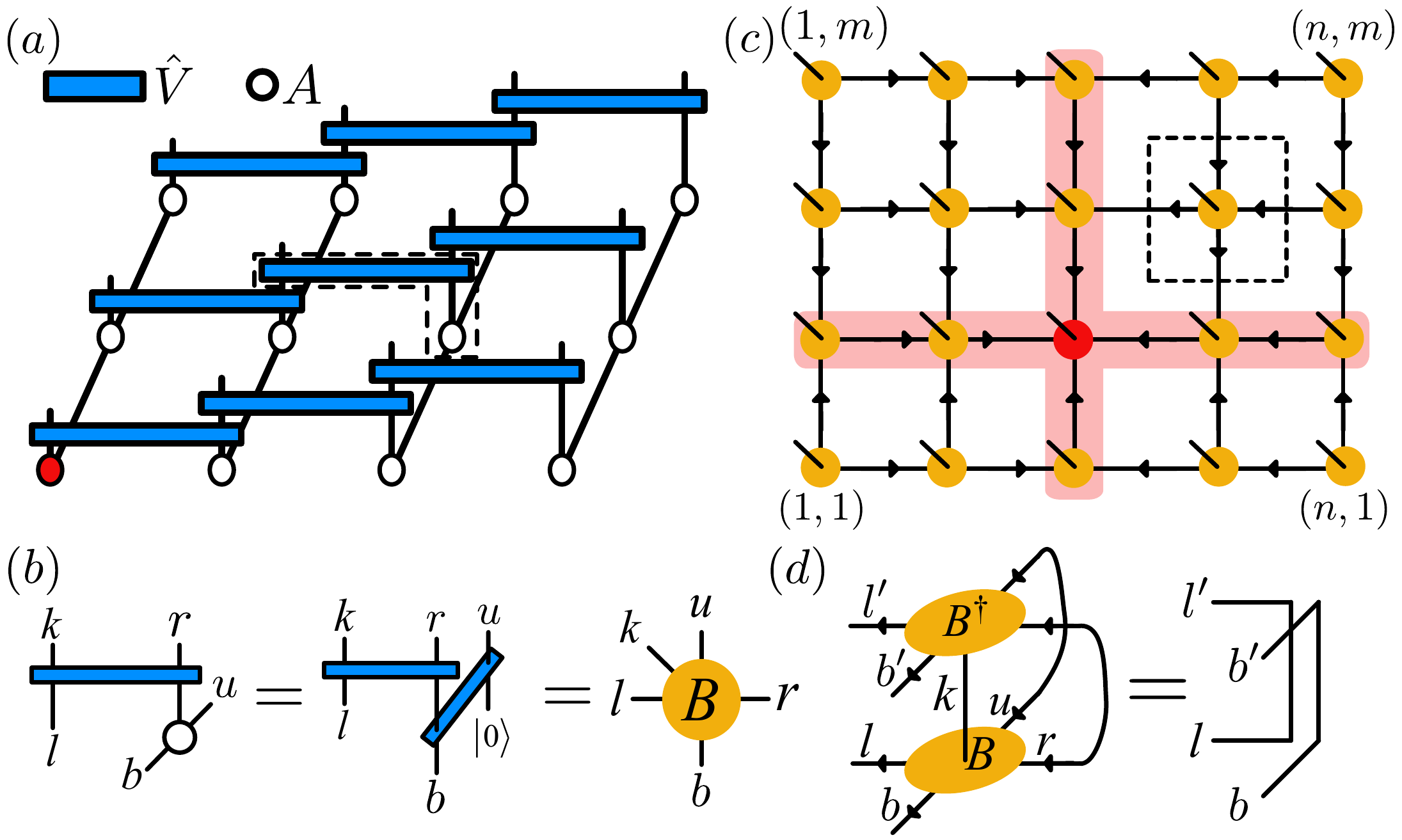}
        \caption{(a) 2D SGS are constructed by first preparing multiple columns of MPS $\{{| \psi _{{\rm{MPS}}}^{{i'}}} \rangle \}$ with tensors $\{A\}$, and then coupling neighboring columns of MPS with linear sequential unitaries $\{ \hat V\}$ of length $L_p$ ($L_p = 2$ here). 
        The red dot denotes the source point of the corresponding P-PEPS.
        (b) The tensor inside the dashed box in (a) can be viewed as two connected two-qudit unitaries and identified as a PEPS tensor.
        (c) Isometric tensor network states (isoTNS).
        The red shaded lines (dot) are the orthogonality hypersurfaces (orthogonality center).
        (d) The isometry condition of the tensor inside the dashed box in (a) and (c).}
        \label{iso_SGS}
\end{figure}

SGS [c.f.~\cref{iso_SGS}(a)] are defined in terms of linear sequential circuits comprising unitaries $\{\hat V_{[i,j]}\}$ of length $L_p$ acting on rows across qudits whose columns have been prepared in MPS~\cite{Banuls2008}
\begin{equation} \label{SGS_def}
    | {\psi _{{\rm{sgs}}}} \rangle  = \prod\limits_{i = 1}^{n-s} {\prod\limits_{j = 1}^{m} {{{\hat V}_{[ {i,j} ]}}} \mathop  \otimes \limits_{i' = 1}^n | {\psi _{{\rm{MPS}}}^{{i'}}} \rangle }.
\end{equation}
The MPS in each column can be put in canonical form, such that they can be written as linear sequential circuits~\cite{Schon2005,Banuls2008}. This allows us to identify the tensor of the corresponding PEPS as two overlapping $L_p$-qudit unitaries [c.f.~\cref{iso_SGS}(b)].
These two unitaries are contained in a $L_p\times L_p$ plaquette unitary. Thus, each SGS can be written as an RP-PEPS, with the source point at the bottom left of the lattice in the case of \cref{iso_SGS}(a).

To be precise, let us denote the class of SGS (P-PEPS) on an $n\times m$ lattice with circuit length (plaquette length) $L_p$ as ${\rm SGS}_{n\times m}^{L_p}$ (${\textrm{P-PEPS}}_{n\times m}^{L_p}$), we have
\begin{equation} \label{state_rel_lsg}
{\rm{SGS}}_{n\times m}^{{L_p}} \subset {\textrm{RP-PEPS}}_{n\times m}^{{L_p}}.
\end{equation}
The tensors of SGS in the bulk satisfy an isometry condition shown in \cref{iso_SGS}(d), 
which is the same condition as is obeyed by the tensors in isoTNS~\cite{Zaletel2020}.
Indeed, as we show in the following, these classes are closely related.

IsoTNS are PEPS [\cref{peps_def}] in which all tensors satisfy isometry conditions that depend on their position in the lattice. Specifically, when all incoming indices of a tensor (denoted by incoming arrows in \cref{iso_SGS}c) and the physical index are contracted with corresponding indices of the complex conjugate of that tensor, the remaining indices yield the identity.
For example, the tensor in the dashed box in \cref{iso_SGS}(c) obeys [c.f.~\cref{iso_SGS}(d)]
\begin{equation} \label{isop}
\sum\limits_{k,ur} {{B_{[ {i,j} ]}^k}_{lurb}{{( {{B_{[ {i,j} ]}^k}_{l'urb'}} )}^*}}  = {\delta _{bb'}}{\delta _{ll'}}.
\end{equation}
The red shaded lines in \cref{iso_SGS}(c) are called \textit{orthogonality hypersurfaces}, which only have incoming arrows, and their intersection is the \textit{orthogonality center} (OC)~\cite{Zaletel2020}. 

\begin{figure}[t]
	\centering
	\includegraphics[width=0.48\textwidth]{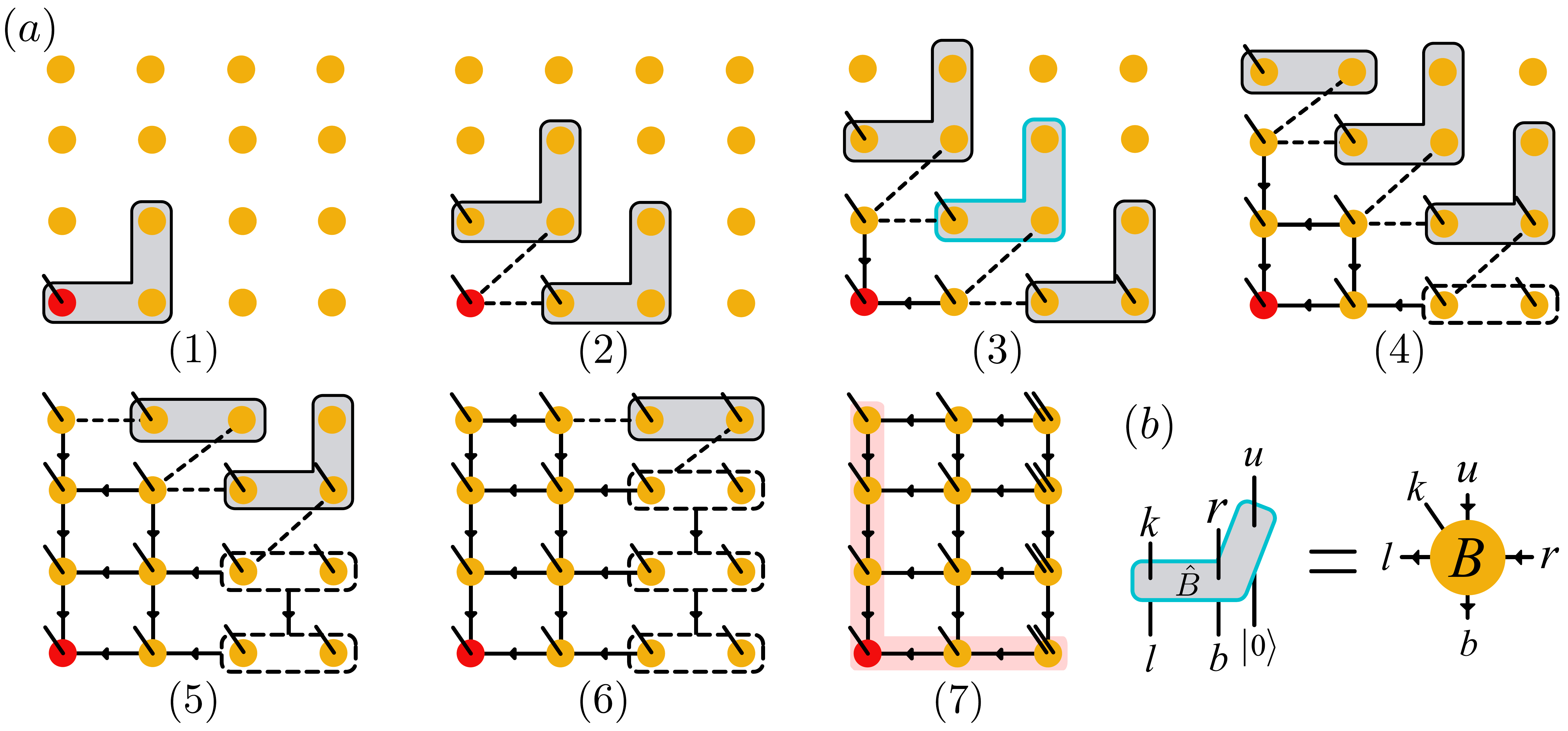}
        \caption{Radial preparation of isoTNS. (a) The protocol starts from an initial state ${\left| 0 \right\rangle ^{ \otimes N}}$, and we sequentially apply three-qudit unitaries that couple neighboring qudits vertically and horizontally following the order in steps 1-6. Each unitary becomes a tensor of the resulting PEPS at the location indicated by the index sticking out. The dashed lines are guiding lines to indicate the connection to previously applied unitary blocks. At the end (4-6) we group the sites of the last two columns (dashed boxes) to satisfy the isometry conditions. This circuit can prepare arbitrary isoTNS [cf.\ \cref{iso_SGS}(c)] of the geometry shown in step 7, with bond dimension $d$, and physical dimension $d$. 
(b) Each three-qudit unitary can be identified as an isoTNS tensor, which satisfies the isometry condition \cref{isop}.}
        \label{seq_iso}
\end{figure}

One can prepare isoTNS as RP-PEPS, but restricting the unitaries in the bulk to be `L'-shaped, as shown in \cref{seq_iso}(a1-6).
The required three-qudit unitaries can be written as
\begin{equation} \label{iso_unit_atom}
    {\hat B_{[ {i,j} ]}} = \sum\limits_{lurb,k} {{B_{[ {i,j} ]}^k}_{lurb}} {| {k,r,u} \rangle }\langle {l,b,0} |,
\end{equation}
where $|k,r,u\rangle \equiv |{k_{\left[ {i,j} \right]}},{r_{\left[ {i,j + 1} \right]}},{u_{\left[ {i + 1,j + 1} \right]}}\rangle$, and the tensor ${{B_{[ {i,j} ]}^k}_{lurb}}$ automatically satisfies the isometry condition \cref{isop}. Thus each unitary creates an isoTNS site [c.f.~\cref{seq_iso}(b)].

Sequentially applying the gates shown in \cref{seq_iso}(a1-6) gives rise to the tensor contraction pattern shown in \cref{seq_iso}(a7), which represents an arbitrary isoTNS with OC in the corner.
The generated isoTNS has bond dimension $d$ and physical dimension $d$, except at the right boundary, where two sites of each row are combined to form a site with physical dimension $d^2$.
Note that arbitrary isoTNS of that geometry with a uniform physical dimension can be embedded in that state by setting the rightmost qudits to zero and treating them as ancillas. Moreover, it is clear from \cref{seq_iso}(a) that the circuit depth for preparing isoTNS is ${T_{{\rm{iso}}}} \approx n + m$.
We show in the SI~\cite{Note1} that: i) by extending the indices of `L'-shaped unitaries to $2s+1$ qudits with $s = \lceil {{{\log }_d}D} \rceil$ and changing the source point of the RP-PEPS, isoTNS with arbitrary bond dimension $D$ and with OC in the bulk can be prepared. ii) This protocol can be generalized to prepare isoTNS of higher dimensions. Therefore, isoTNS on arbitrary lattices (of size ${n_1} \times ... \times {n_q}$) admit \textit{exact} representations as sequential quantum circuits, with the circuit depth scaling as 
\begin{equation} \label{dep_iso}
{T_{{\rm{iso}}}} \approx \sum\limits_{i = 1}^q {{n_i}}.
\end{equation}

The above observation shows that isoTNS $\subset$ RP-PEPS. A similar relation also holds between P-PEPS and F-PEPS~\cite{Pichler2017}, here understood as a generalization to qudits and with arbitrary photon feedback. F-PEPS can be viewed as isoTNS on a lattice with different connectivity~\cite{Pichler2017}.
If we denote ${\rm{isoTNS}}_{n\times m}^{D,d}$ (${\textrm{F-PEPS}}_{n \times m}^{D,d}$) as the class of isoTNS (F-PEPS) on a $n\times m$  lattice with bond dimension $D$ and physical dimension $d$, we prove in the SI~\cite{Note1} that isoTNS (F-PEPS) is contained in RP-PEPS (P-PEPS) with a slightly larger lattice
\begin{equation} \label{rel_iso}
{\rm{isoTNS}}_{n \times m}^{D,d} \subset {\textrm{RP-PEPS}}_{(n + 2s) \times (m+2s)}^{2s + 1},
\end{equation}
\begin{equation} \label{rel_fp}
{\textrm{F-PEPS}}_{n \times m}^{D,d} \subset {\textrm{P-PEPS}}_{(n + s) \times (m+s)}^{2s + 1}.
\end{equation}

Having established that both SGS and isoTNS are RP-PEPS with `L'-shaped unitaries, we note that SGS has a further condition on the unitaries, namely that they can be decomposed into two unitaries corresponding to the two arms of the L (see \cref{iso_SGS}b), which indicates that SGS are a subclass of isoTNS. 
This has direct consequences for the states. While in SGS, local observables can efficiently be calculated anywhere in the lattice, in isoTNS, this requires shifting the OC, which can only be done approximately. Their precise relation is~\cite{Note1}
\begin{equation} \label{state_rel_lsg}
{\rm{SGS}}_{n \times m}^{{L_p}} \subset {\rm{isoTNS}}_{n \times m}^{{d^{L_p(L_p-1)}},d}.
\end{equation}

Finally, we note that, since the isometry of the PEPS tensors derives directly from the unitarity of the preparation circuit, we further show that RP-PEPS can be expressed as isoTNS on lattices with unusual connectivities~\cite{Note1}.

\textit{Generating RP-PEPS of flying qub(d)its.---} Ref.~\cite{Schon2005} introduces a protocol to prepare arbitrary photonic MPS of bond dimension $D$ and physical dimension $d$ using a photon source comprising a $D$-level ancilla $A_1$ and a $d$-level emitter $E_1$. The MPS is prepared by repeatedly applying a unitary on the joint ancilla-emitter system, followed by swapping the emitter state into a flying photon, defined in terms of the photon emission isometry $M_{\rm ph}$
\begin{equation} \label{ap_map}
    {M_{\rm ph}}:{| k \rangle _d} \to {| 0 \rangle _d}{| k \rangle _{{\rm{\rm ph}}}},\qquad k \in ( {0,...,d - 1} ).
\end{equation}
 \begin{figure}[tb]
	\centering
	\includegraphics[width=0.48\textwidth]{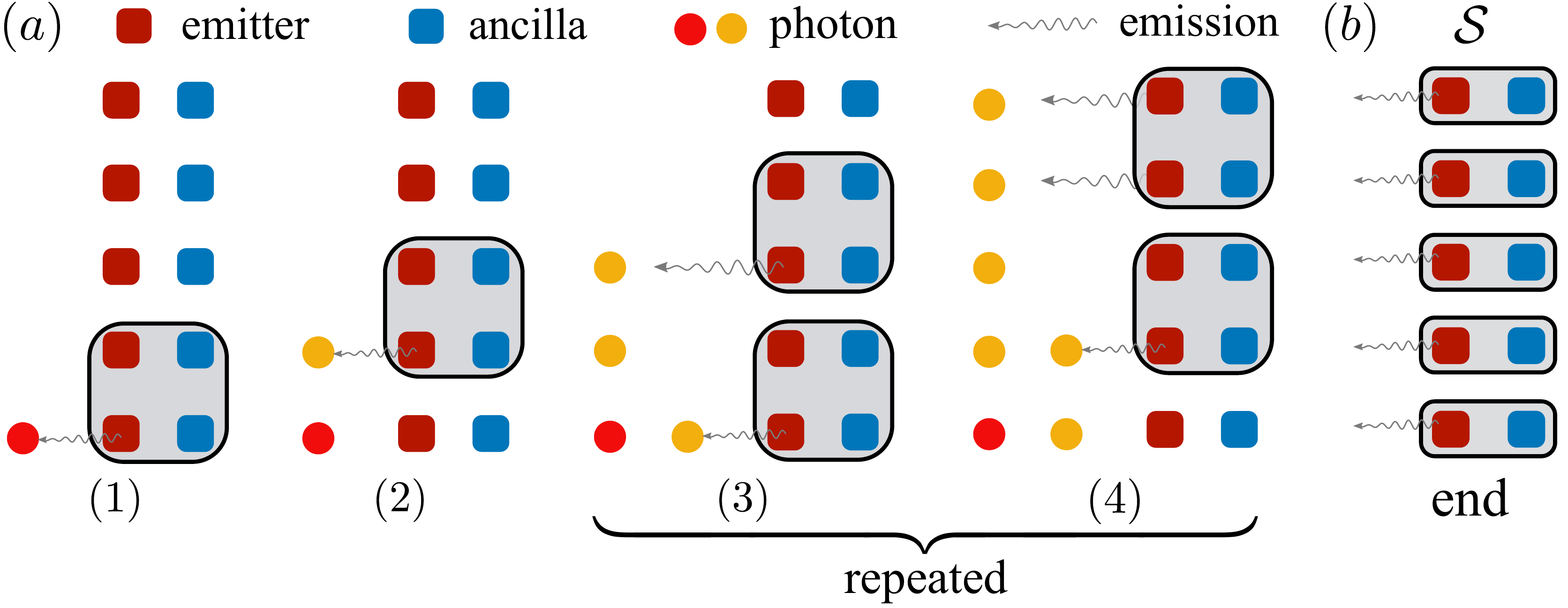}
        \caption{Generation of photonic RP-PEPS using an array of coupled sequential photon sources. (a) In the preparation of the $[i,j]$-th RP-PEPS site, we apply a unitary ${\hat U_{[ {i,j} ]}}$ followed by a photon emission $M^{i}_{\rm ph}$ of the emitter $E_j$. After the initial steps (1) and (2), steps (3) and (4) will be repeated. (b) At the end of the protocol, we swap the excitations on the ancillas to the emitters and then convert them to photons, denoted as $\cal S$.}
                \label{isop_gen}
\end{figure}
Now we extend the above protocol by considering an array of $m$ sequential photon sources coupled to each other as shown in \cref{isop_gen}, and show that photonic RP-PEPS can be prepared with this setup.

The protocol is shown in \cref{isop_gen}. Here we assume each ancilla has a dimension $D=d^{L_p-1}$, so it can be thought of as $L_p-1$ qudits. Starting with the ancillas $\{A_j \}$ and emitters $\{E_j\}$ in their ground state $| {{\varphi _0}} \rangle  = | {{0_{\{ {{A_j}} \}}}} \rangle  \otimes | {{0_{\{ {{E_j}} \}}}} \rangle $, in the step to prepare the $(i,j)$-th site of the RP-PEPS, first we apply a unitary ${\hat U_{[ {i,j} ]}}$ that acts on the ancillas $\{A_j,...,A_{j+L_p-1}\}$ and emitters $\{E_j,...,E_{j+L_p-1}\}$ [see \cref{isop_gen}(a) for $L_p=2$ case]. After the unitary, we trigger the photon emission from the emitter $E_j$ (denoted as $M^{j}_{{\rm{ph}}}$ [cf.~\cref{ap_map}]). To disentangle the ancilla from the photons, in the last $L_p-1$ steps of the protocol, we sequentially swap the effective $L_p-1$ qudits contained in the ancilla into photons (c.f.~\cref{isop_gen}b), an operation collectively denoted by $\cal S$.
The final state of the system is $\left| {{\varphi _0}} \right\rangle {\left| {{\psi _{\rm rp}}} \right\rangle _{\rm ph}}$, with the photonic state
\begin{equation} \label{P-PEPS_ph}
|\psi _{{\rm{rp}}}{\rangle _{{\rm{ph}}}} = \left\langle {{\varphi _0}} \right|{\cal S}\prod\limits_{i = 1}^{n - {L_p} + 1} {\prod\limits_{j = 1}^{m - {L_p} + 1} {(M_{{\rm{ph}}}^j{{\hat U}_{[i,j]}})} \left| {{\varphi _0}} \right\rangle }.
\end{equation}
$|\psi _{{\rm{rp}}}{\rangle _{{\rm{ph}}}}$ is an arbitrary RP-PEPS with OBC and plaquette size $L_p$, with its source point at the first photonic qudit. The circuit depth is the same as that of matter-based lattice case [c.f.~\cref{radial_speed}]. The same protocol also allows to prepare isoTNS~\cite{Note1}, and this setup can be used to prepare photonic F-PEPS with circuit depth $O(N)$~\cite{Note1}.

Notice that, at the boundary of the photon source array, the photon emission process emits multiple photons, as visualized in \cref{isop_gen}(a4). In contrast to the protocol that produces RP-PEPS on a matter-based lattice, here the overlapping of gates along the horizontal direction are produced by acting on the ancillas. In Ref.~\cite{zycq} we propose a cavity-transmon setup to realize of this protocol, and elaborate on how to create the two-dimensional photonic cluster state and the toric code state~\footnote{Note that for the cluster state generation, there are specific schemes \cite{lindner2009,Economou2010} that do not need ancillas and SWAP gates. These schemes thus require less resources than the present scheme.}.

\emph{Conclusion.---}We have introduced P-PEPS and its subclass RP-PEPS, which constitute a natural generalization of sequential preparation protocols from one to higher dimensions. These states satisfy area-law entanglement by construction, combine the capacity to host long-range correlations, topologically ordered states, and a large subclass of PEPS with a simple and efficient preparation protocol.
Our work helps to clarify the relation between various relevant classes of PEPS, including SGS~\cite{Banuls2008}, F-PEPS ~\cite{Pichler2017} and isoTNS~\cite{Zaletel2020}, that we show SGS $\subset$ isoTNS $\subset$ RP-PEPS, and F-PEPS $\subset$ P-PEPS.

The family of states we introduce come with explicit protocols that prepare them in matter-based and photon-based lattices, which makes them promising targets for near-term experimental realization. Furthermore, one can include several layers of sequential plaquettes to increase the expressivity of the ansatz.

\hspace*{\fill} \\

We acknowledge funding from ERC Advanced Grant QUENOCOBA under the EU Horizon 2020 program (Grant Agreement No. 742102), and within the D-A-CH Lead-Agency Agreement through project No. 414325145 (BEYOND C), and the European Union's Horizon 2020 research and innovation program under Grant No. 899354 (FET Open SuperQuLAN).

\nocite{Tepaske2020,Raussendorf2006}

\newpage

\appendix

{\huge Supplementary Materials}
\tableofcontents




\section{Properties of P-PEPS}

\subsection{Bond dimension of the PEPS representation of P-PEPS}

As discussed in the main text, one can obtain the PEPS representation of P-PEPS by decomposing the plaquette unitaries into projected entangled-pair operators (PEPO)~\cite{Cirac2020} (shown in Fig.~1d in the main text). To be precise, given a plaquette unitary of side length $L_p$, the resulting bond dimension of each PEPO is trivially bounded by $O(d^{L_p^2})$, which come from the MPO decomposition of given unitary acting on $L_p^2$ qudits. Moreover, there are $L_p^2$ plaquette unitaries acting on each single site. Thus the bond dimension of the PEPS representation of P-PEPS is bounded from above by $D \le O(d^{L_p^4})$.

\subsection{Efficient computation of certain expectation values}

Let us assume we have a local operator $\hat O$. Then its expectation value in a given P-PEPS is
\begin{equation} \label{obs_calc}
    \left\langle {{\psi _{\rm p}}} \right|\hat O\left| {{\psi _{\rm p}}} \right\rangle  = {\left\langle 0 \right|^{ \otimes N}}\hat U_{{{\vec v}_1}}^\dag ...\hat U_{{{\vec v}_N}}^\dag {\hat O}{{\hat U}_{{{\vec v}_N}}}...{{\hat U}_{{{\vec v}_1}}}{\left| 0 \right\rangle ^{ \otimes N}}.
\end{equation}
Depending on the ordering $\cal P$ of the $| {{\psi _{\rm p}}} \rangle$ and the location of $\hat O$ in the lattice, it is possible to directly cancel some gates in \cref{obs_calc} with their hermitian conjugate. 

\begin{figure}[h!]

	\centering
	\includegraphics[width=0.48\textwidth]{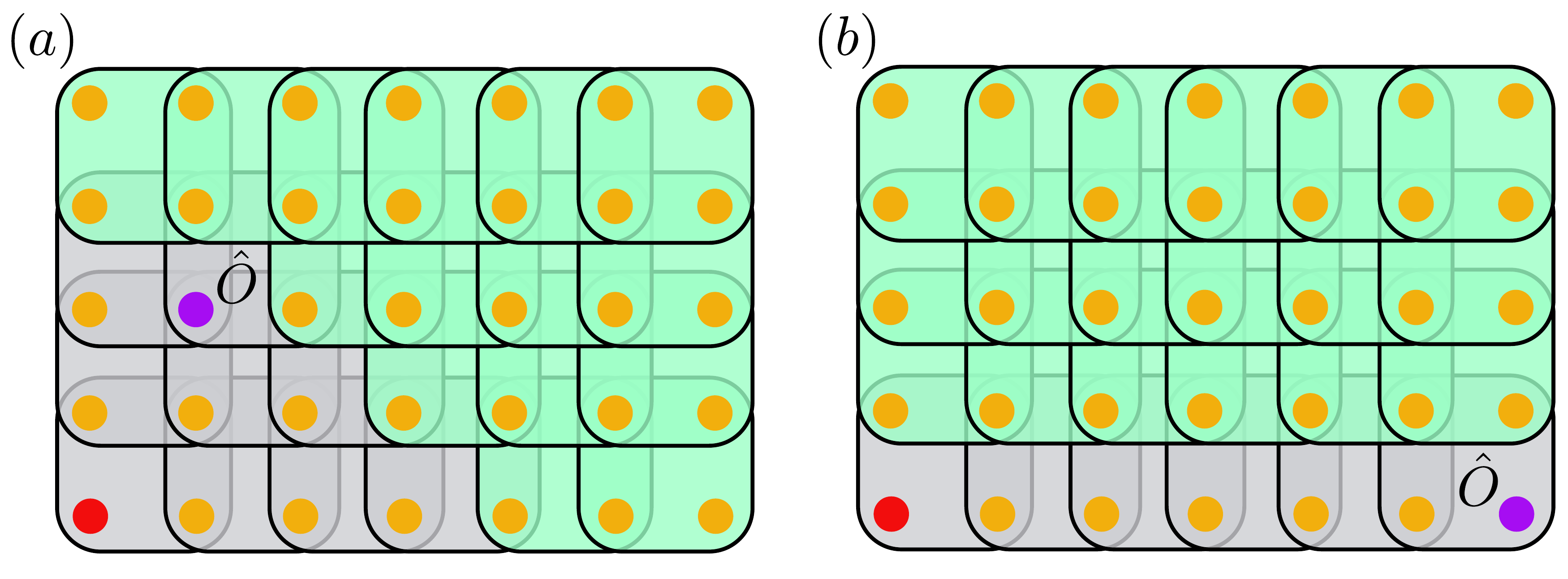}
        \caption{Efficient computation of certain expectation values for RP-PEPS ($L_p=2$ here). Here the red dot marks the source point. When we want to compute the expectation values of an operator $\hat O$ located at the site marked by the purple dot, that is (a) geometrically close to the source point, or (b) along the line that crosses the source point and extends along the preferred direction, a majority of unitaries in Eq.~(2) in the main text will be canceled with their hermitian conjugates, marked by the green plaquettes here.}
        \label{eff_comp}
\end{figure}

We illustrate this effect in \cref{eff_comp} using a RP-PEPS with the source point in a corner of the lattice. For RP-PEPS, the ordering $\cal P$ reflects the distance of the unitary to the source point. In this case, the expectation values of local operators that are geometrically close to the source point can be computed efficiently [c.f.~\cref{eff_comp}(a)]. Moreover, due to the choice of the preferred direction for plaquettes at equal distance, one finds that also along the line starting from the source point and along the preferred direction, expectation values can be computed efficiently [c.f.~\cref{eff_comp}(b)].

Note that this is very similar to what happens in isometric tensor networks, in which expectation values of operators close to the OC or the orthogonal hypersurfaces can be calculated efficiently~\cite{Zaletel2020}.

\subsection{Long-range correlation in the P-PEPS, and comparison to brickwall circuits}
\begin{figure}[h!]
	\centering
	\includegraphics[width=0.48\textwidth]{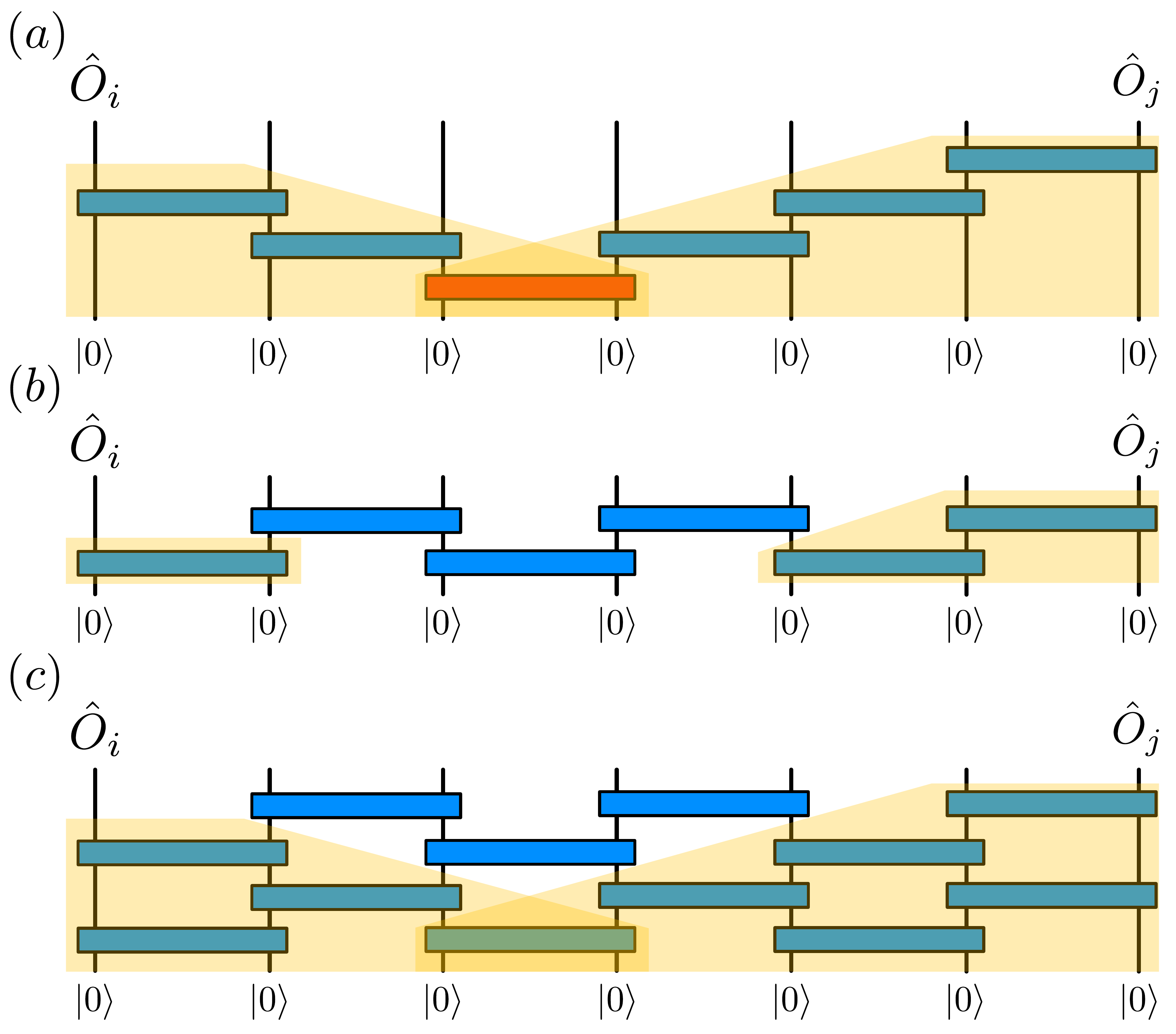}
        \caption{The correlation in sequential circuit pattern for P-PEPS and the brickwall circuits in the one-dimensional case ($L_p=2$ here). (a) The sequential circuit pattern ensures long-range correlation, as the reversed light cone (the yellow regions) always have overlap, at least at the position of the first unitary (marked by the red box). (b) Given the same amount of gates $O(N)$ for the sequential circuit pattern, the brickwall circuit can only create short-range correlations. (c) To also create long-range correlation using brickwall circuit, the circuit depth scale the same as the circuit pattern for RP-PEPS $T_{\rm rp}$ [Eq.~(1) in the main text], which also mean a $T_{\rm rp}$-fold increase on the number of gates needed, compared to the sequential circuit in panel (a).}
        \label{psg_corr}
\end{figure}

In case that the plaquette unitaries for creating P-PEPS [Eq.~(2) in the main text] overlap with earlier ones, the resulting P-PEPS generally have long-range correlations among arbitrary locations in the lattice. To illustrate this, first let us compare such one-dimensional P-PEPS and the brickwall circuit~\cite{Gopalakrishnan2019}, shown in \cref{psg_corr}. Here the `plaquette' gates for one-dimensional P-PEPS are local gates of $L_p$ qub(d)its.

In this section, we assume the shape of unitaries in the brickwall circuits is the same as in the case of P-PEPS. A similar discussion between these two circuit structures can be found in Ref.~\cite{lin2021real}.

Given a state prepared by quantum circuits, the correlations between two locations are non-zero if the reversed `light cones' starting from these two locations overlap. This is illustrated in one dimension in \cref{psg_corr}(a).
If we take the same number of unitaries and arrange them in a brickwall circuit (this also belong to P-PEPS), it will be a finite-depth circuit of depth $O(L_p)$ [shown in \cref{psg_corr}(b) for $L_p=2$], which only creates correlations up to a distance $O(L_p)$.
To create long-range correlations in brickwall circuits requires a circuit depth of at least $O(n/L_p)$ [c.f.~\cref{psg_corr}(c)], which uses many more gates.

The above observation also holds for higher dimensional P-PEPS. Thus, when the plaquette unitaries in Eq.~(2) in the main text have overlap with earlier ones, the reversed light cones of two arbitrary locations in the lattice always overlap, at the very least on the first plaquette.
For a $N = {n_1} \times ... \times {n_q}$ lattice, it generally take a brickwall circuit of depth $O(\mathop {\max }\limits_i {n_i})$ to produce long-range correlations among the whole lattice, thus takes $O(N\cdot \mathop {\max }\limits_i {n_i})$ gates, many more than for P-PEPS preparation.

Thus P-PEPS contain states with long-range correlation among arbitrary locations in the lattice. Moreover, such long-range correlated states in the form of RP-PEPS can be prepared particularly efficiently, with circuit depth scale as the maximum edge length of the lattice [Eq.~(1) in the main text].

\section{Preparing arbitrary 2D isoTNS on lattices of stationary qub(d)its}

In this section, we show how to extend the circuit pattern in Fig.~3(a) in the main text to prepare arbitrary 2D isoTNS in a square lattice.

\subsection{Increase the isoTNS bond dimension}

\begin{figure}[h!]
	\centering
	\includegraphics[width=0.48\textwidth]{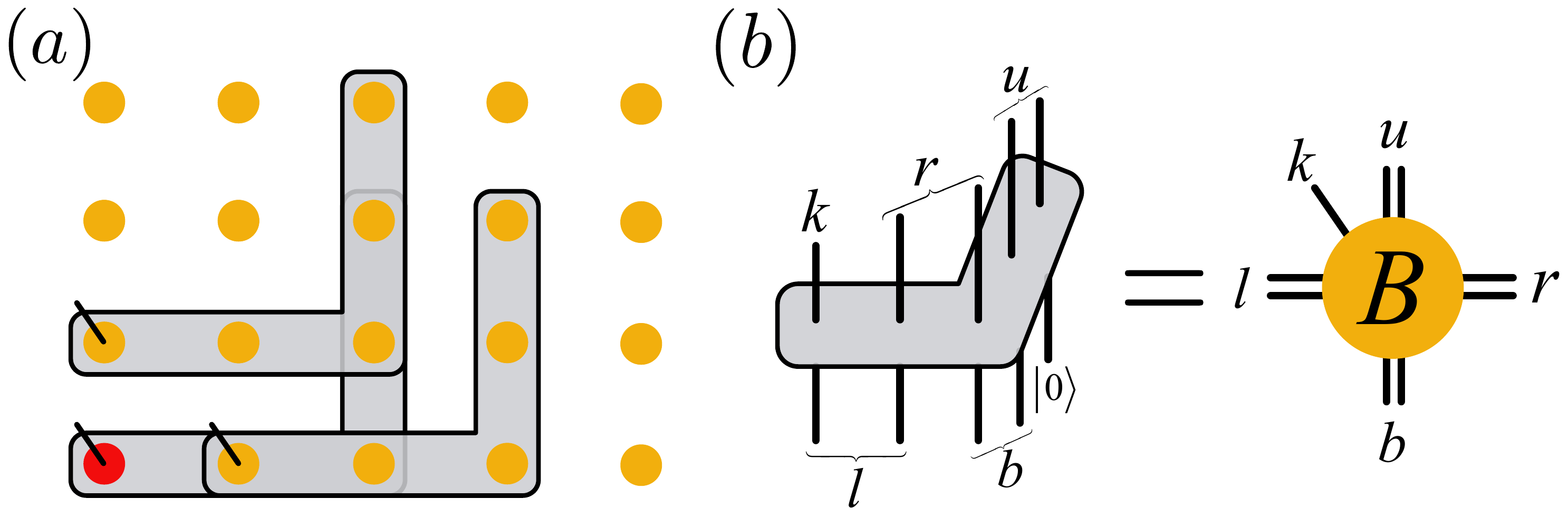}
        \caption{(a) One can prepare isoTNS with higher bond dimension by applying gates that have more common sites. (b) The applied gates converted to an isoTNS with higher bond dimension. Here each black line is a bond of dimension $d$, thus the virtual bond dimension here is $D=d^2$.}
        \label{bond_incr}
\end{figure}

One can increase the bond dimension of isoTNS prepared by sequential circuit in Fig.~3(a) in the main text by making the gates acting on more common sites. In particular, the gate $\hat B_{[i,j]}$ that generate isoTNS of bond dimension $D$ acts on the site $\left\{ {\left( {i,j} \right),...,\left( {i,j + s} \right),\left( {i + s,j + s} \right)} \right\}$ where $s = \left\lceil {{{\log }_d}D} \right\rceil$. The example of $s=2$ is shown in \cref{bond_incr}(a), with the identification of the PEPS bonds shown in \cref{bond_incr}(b). 

Notice that, unlike the case of RP-PEPS [Eq.~(1) in the main text], increasing the indices of the `L'-shape unitaries does not change the circuit depth Eq.~(8) in the main text of preparing isoTNS, since it is clear from \cref{bond_incr}(a) that we can always parallelize the two unitaries on the upper and right side of a given `L'-shape unitary.

\subsection{Preparing isoTNS with OC at arbitrary location}

\begin{figure}[h!]

	\centering
	\includegraphics[width=0.48\textwidth]{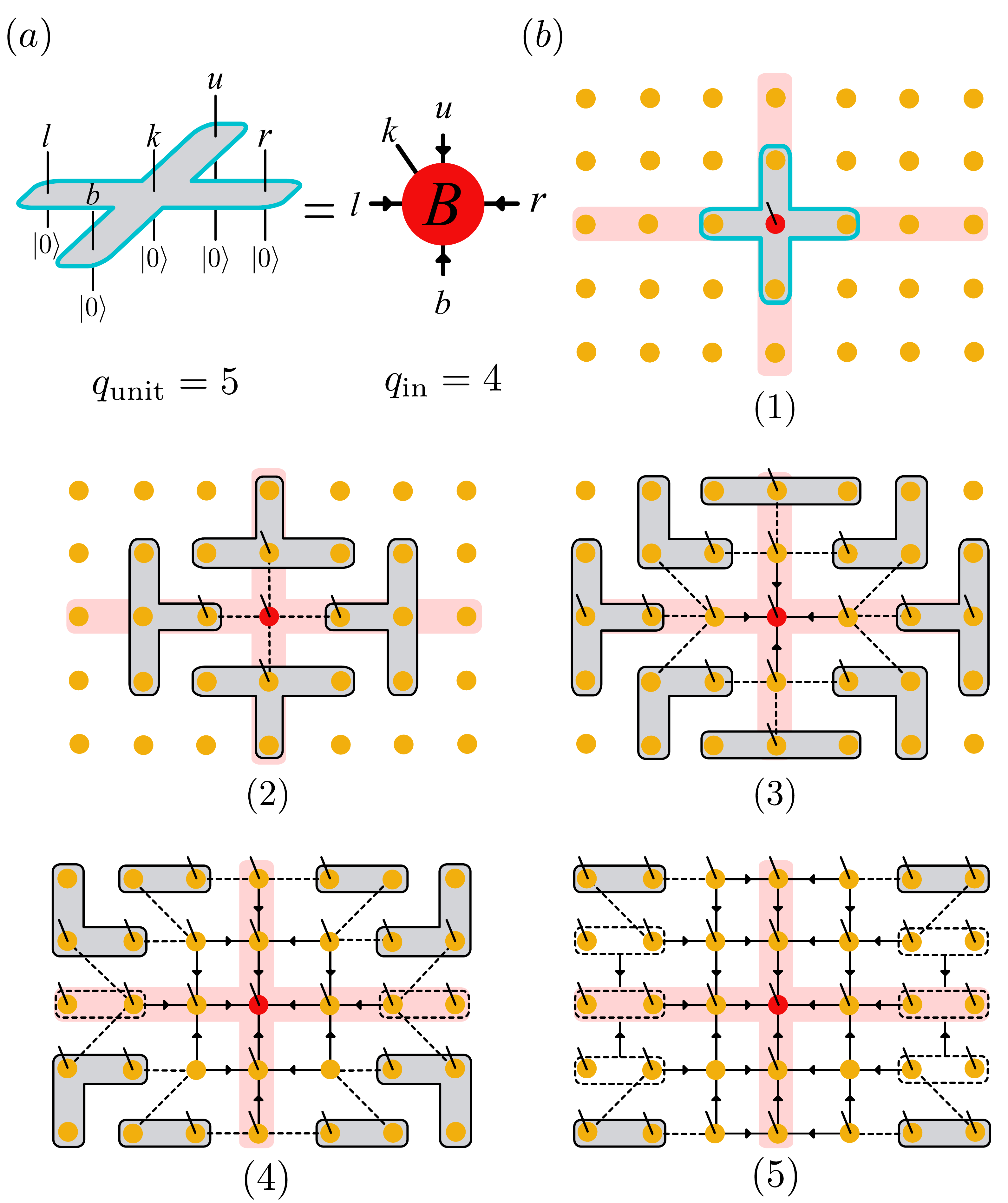}
        \caption{The preparation of the isoTNS with OC in the bulk. (a) The tensor of isoTNS at OC has $q_{\rm in}=4$ incoming virtual bonds, which correspond to a $q_{\rm unit}=5$ qudit unitary acts on the initial state, which follows the general relation \cref{real_in_unit}. (b) The preparation procedure. Here we use the same notations as in Fig.~3(a) in the main text. One first prepare the OC in step 1, and then sequentially grow the region along all directions. The tensors in the orthogonality hypersurface (denoted by the red shaded areas) generally have $q_{\rm in}=3$ incoming virtual bonds, thus are prepared by four-qudit unitaries shown in step 2 and 3. After that, one can use the same procedure as in Fig.~3(a) in the main text to grow the four regions of the lattice, eventually, prepare an isoTNS with OC in the bulk.}
        \label{iso_mix}
\end{figure}

The protocol in Fig.~3(a) in the main text prepares arbitrary isoTNS of the geometry shown in step 7 of Fig.~3(a) in the main text, where the OC is at the corner of the lattice, marked with the red dot. In general, the number of incoming virtual bonds $q_{\rm in}$ in an isoTNS tensor is directly related to the number of outgoing indices $q_{\rm unit}$ for the unitary, as
\begin{equation} \label{real_in_unit}
	q_{\rm in} = q_{\rm unit} -1.
\end{equation}
For example, the isoTNS tensor shown in Fig.~3(b) in the main text are created with three-qudit unitaries ($q_{\rm unit}=3$), since there are two incoming indices ($q_{\rm in}=2$) for that tensors. When OC at an arbitrary location inside the bulk (like that in Fig.~2a in the main text), the tensor at OC has $q_{\rm in}=4$ incoming virtual bonds, thus one need a 5-qudit unitary to prepare it, which is shown in \cref{iso_mix}(a). Thus we can use the circuit pattern in \cref{iso_mix}(b) to prepare isoTNS with OC in the bulk of the lattice. In step 1, we prepare the tensor at OC by applying a 5-qudit unitary. Then the tensors at the orthogonality hypersurface are prepared by 4-qudit unitaries, for example, shown in step 2. Then one can apply the same procedure as that in Fig.~3(a) in the main text to grow the size of the prepared isoTNS on four regions separated by the orthogonality hypersurface, and use the sites on both left and right boundaries of the lattice as ancillas, to create an arbitrary isoTNS of bond dimension $d$ and physical dimension $d$, with OC at an arbitrary location of the lattice. The method to increase the bond dimensions also naturally applies in this scenario.

Preparing isoTNS with OC inside the bulk of the lattice not only makes the class of isoTNS with open boundary conditions we prepare complete but also has practical relevance for the state preparation speed. As the region of prepared sites expands along all directions simultaneously, one can prepare an isoTNS with OC at the center four times faster than one with the OC at the corner. We also point out that the isoTNS with OC inside the bulk is also naturally contained in the class of P-PEPS, as we can just cover the local gates shown in \cref{iso_mix}(b) inside appropriate plaquettes and the ordering of plaquette gates naturally captures the sequential ordering in \cref{iso_mix}(b).

\subsection{Ancilla in preparing isoTNS with uniform bond dimensions}

As shown in the main text Fig.~3(a) and the previous sections [c.f.~\cref{iso_mix}(b) and \cref{fp_circuit}(b)], one need to use some ancillas when preparing isoTNS. This comes from the mechanism of connecting the tensors of the isoTNS, that we need to let the neighboring gates have overlapping regions to create virtual bonds of the tensors. As a result, when reaching the boundary of the system, we have to merge sites to obtain a general isoTNS of that geometry, for example as shown in Fig.~3(a6-7) in the main text. Without merging, the resulting state would lack vertical bonds between the rightmost sites. Clearly one can embed arbitrary isoTNS of uniform bond dimension and smaller lattice into that geometry, treating the sites on the boundary as ancillas. 

\begin{figure}[h!]
	\centering
	\includegraphics[width=0.48\textwidth]{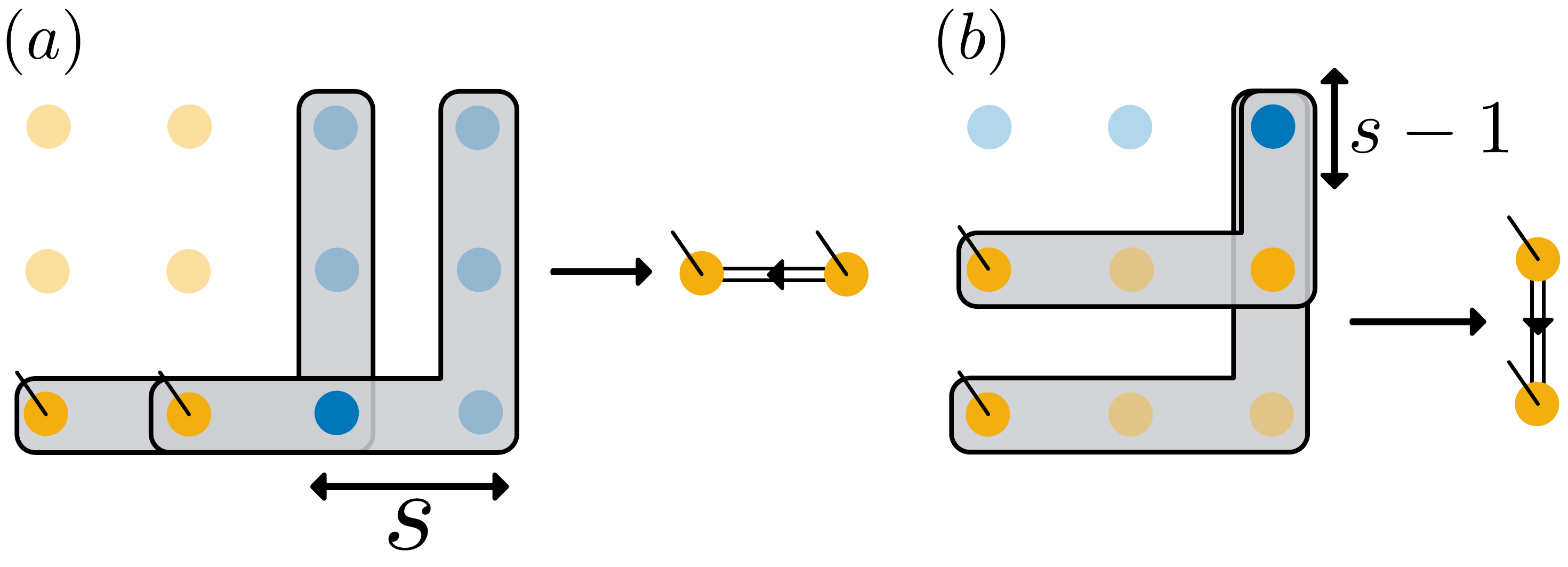}
        \caption{Using ancillas to create virtual bonds of isoTNS with uniform bond dimension. Here the blue dots are the sites used as ancillas. The physical site of the tensor and the overlapping sites of the two gates are marked more prominently. Panel (a) shows that for the left/right boundary, we need $s$ columns of ancillas, and panel (b) shows for the upper/bottom boundary, one needs $s-1$ rows of ancillas.}
        \label{iso_anci}
\end{figure}

We show the usage of ancilla more explicitly on the right and upper boundary of the lattice in \cref{iso_anci}. For `L'-shape gates of the direction shown there, we need additional $s$ columns and $s-1$ rows of qudits as ancilla, where $s=\left\lceil {{{\log }_d}D} \right\rceil $ for creating isoTNS of bond dimension $D$. Thus overall, given such an isoTNS with OC in the bulk of size $n\times m$, we will need a $(n+2s)\times (m+2s-2)$ lattice, where the sites on the boundary of the lattice act as ancillas.

Moreover, one also needs to use ancilla when creating photonic isoTNS with uniform bond dimensions. We will elaborate it in the photonic isoTNS generation protocol around \cref{isop_si_gen}.

\section{Proof of isoTNS $\subset$ RP-PEPS [Eq.~(9) in the main text]}

It is clear from the previous section that, to produce isoTNS of bond dimension $D$ in a $n\times m$ lattice, one needs a sequential quantum circuit that acts on a $(n+2s) \times (m+2s-2)$ lattice [c.f.~\cref{iso_anci}], where the sites on the boundary are used as ancillas. Also notice that in \cref{iso_anci}(b) the length of the vertical arm of the `L'-shape unitary at the upper boundary of the lattice is also slightly reduced. Thus we need to formally put an additional row of ancillas in \cref{iso_anci}(b) to make that `L'-shape unitary properly covered in a plaquette unitary.

Overall, We can cover the isoTNS preparation circuit into the RP-PEPS preparation circuit on a $(n+2s) \times (m+2s)$ lattice, where the largest plaquette size is determined by the gate that prepares the OC of isoTNS [c.f.~\cref{iso_mix}(a)], where we need a plaquette of side length $L_p = 2s+1$. This thus proves the relation between isoTNS and RP-PEPS [Eq.~(9) in the main text]:
\begin{equation*}
\begin{array}{l}
{\rm{isoTNS}}_{n \times m}^{D,d} \subset {\textrm{RP-PEPS}}_{(n + 2s) \times (m+2s)}^{2s + 1},
\end{array}
\end{equation*}
A similar argument also holds in higher dimensions, that arbitrary isoTNS is contained in RP-PEPS of slightly bigger lattices.

At last, we point out that, if we are allowed to move the location of the ancilla, we could reduce the number of ancillas to the order of $O(\left\lceil {{{\log }_d}D} \right\rceil)$ by efficiently reuse the ancilla that has been disentangled after providing the virtual bonds. For example, in Fig.~3(a) in the main text there are always only two ancillas that are under operation or being entangled to the system during the state preparation procedure.

\section{Representing RP-PEPS as isoTNS}
\label{P-PEPS_iso_sec}
\begin{figure}[h!]
	\centering
	\includegraphics[width=0.48\textwidth]{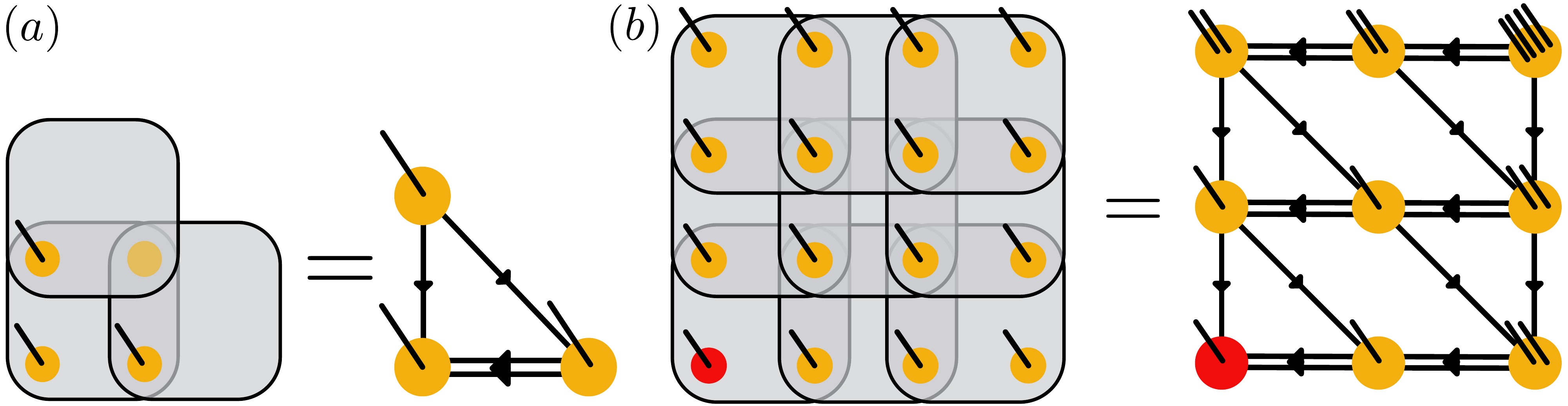}
        \caption{Writing RP-PEPS as an isoTNS. (a) States produced by sequential quantum circuits can be converted to isoTNS, where we identify each unitary as a tensor site. The physical index of the tensor corresponds to the outgoing index(s) that does not connect to other gates, and the virtual indices of the tensors are produced by the connected indices of the gates. (b) The two-dimensional RP-PEPS with grid length $L_p=2$ can be written as an isoTNS with connectivity shown on the right side of the figure. Here each line is a bond of dimension $d$.}
        \label{P-PEPS_iso}
\end{figure}

The sequential application of local quantum gates that have overlapping sites naturally defines a `causal' structure, and the unitarity of gates implies certain isometry conditions. One can view each unitary as a tensor (see \cref{P-PEPS_iso}(a)), with its incoming and outgoing indices that have overlap with other unitaries as the virtual bonds, and the outgoing indices that do not further connect to other gates as the physical bond. In this way, we can write an arbitrary RP-PEPS as isoTNS. The simplest example of $L_p = 2$ RP-PEPS is shown in \cref{P-PEPS_iso}(b). Compare to standard isoTNS~\cite{Zaletel2020} [c.f.~Fig.~2(a) in the main text], the isoTNS representation of RP-PEPS has unusual connectivity and spatial non-uniform bond dimensions and physical dimensions. By further increasing $L_p$ and changing the ordering $\cal P$, the isoTNS representation of RP-PEPS will contain virtual bonds that connect more distant sites and more unusual connectivity.

\section{Preparing higher-dimensional P-PEPS and isoTNS}

It is desirable to generalize our protocols to prepare higher-dimensional P-PEPS and isoTNS, since these states potentially efficiently characterize the ground state of higher-dimensional local Hamiltonians \cite{Tepaske2020}, with prominent examples like the three-dimensional cluster state~\cite{Raussendorf2006,Russo2019}.

Here we briefly discuss how to generalize our scheme to higher dimensions. First, the P-PEPS can naturally be extended to higher dimensions by sequentially applying higher-dimensional cubes of side length $L_p$. For a $q$-dimensional lattice of size $N = {n_1} \times ... \times {n_q}$, the P-PEPS is still of form [Eq.~(2) in the main text]:
\begin{equation*}
\left| {{\psi _{{\rm{p}}}}} \right\rangle  = \prod\limits_{\mu = 1}^N {{{\hat U}_{\vec v_\mu}}} {\left| 0 \right\rangle ^{ \otimes N}},
\end{equation*}
where now the position vector ${{\vec v}_\mu } \equiv ( {i_1^\mu ,i_2^\mu ,...,i_q^\mu } )$. Other quantities like the ordering $\cal P$ and the state with open boundary conditions are defined in the same way as two-dimensional case, and it is clear that the worst-case circuit depth is of $O(N)$.

\begin{figure}[h!]
	\centering
	\includegraphics[width=0.48\textwidth]{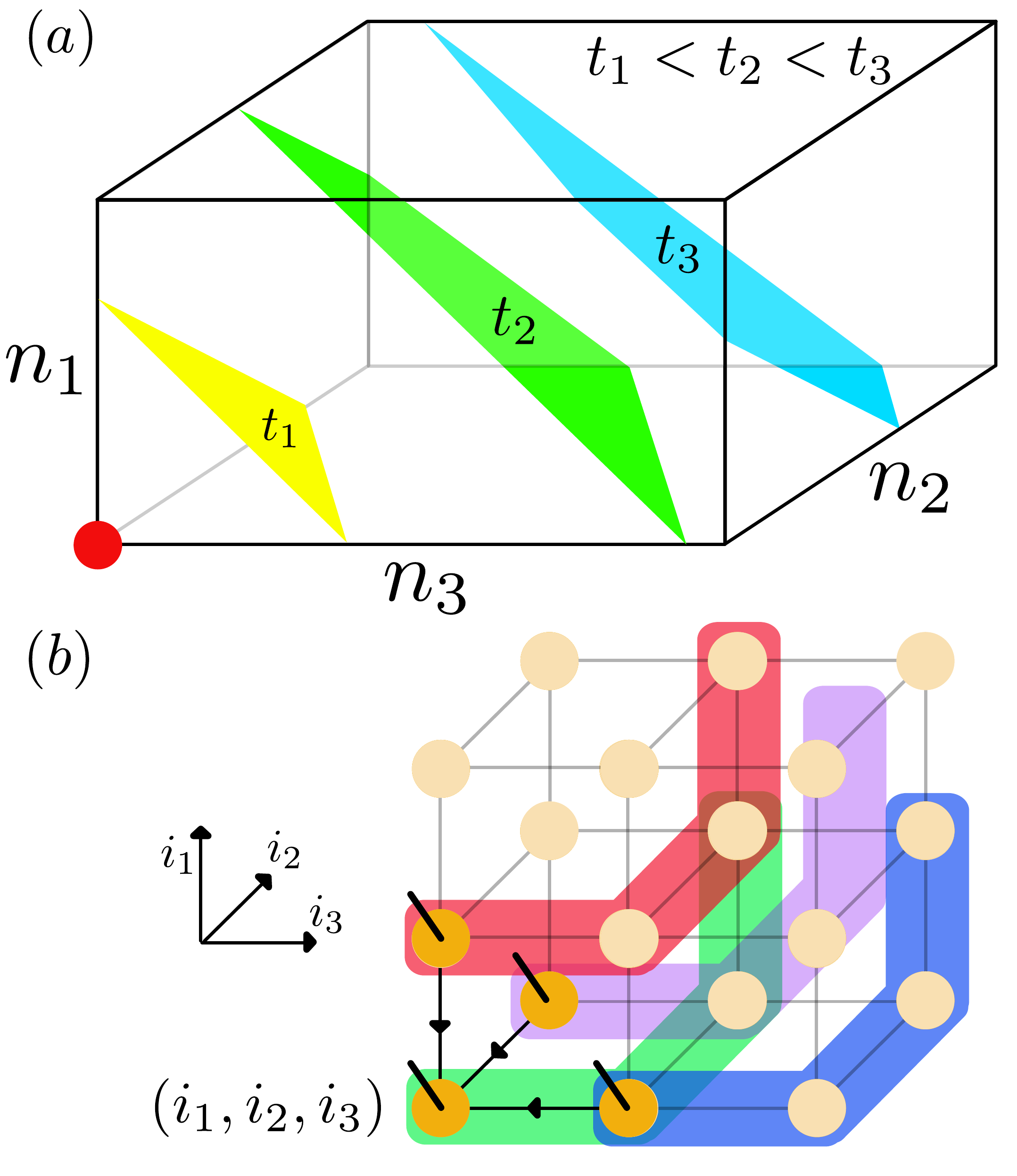}
        \caption{(a) The illustration of generation process of three-dimensional RP-PEPS and isoTNS. Similar to the two-dimensional case [c.f.~Fig.~1(e) in the main text], start from the source point (the red dot), the gate-acted region ballistically expands along all directions, with its boundary at different times denoted by the color surfaces. For RP-PEPS it takes a circuit depth ${T_{{\rm{rp}}}} \approx {n_1} + {L_p}{n_2} + L_p^2{n_3}$ to finish the state preparation [here assume the preferred directions is $(1,2,3)$], and for isoTNS it takes a circuit depth ${T_{{\rm{iso}}}} \approx {n_1} + {n_2} + {n_3}$.
        (b) The unitaries ${\hat B_{[ {{i_1},{i_2},{i_3}}]}}$ (green shades), ${\hat B_{[ {{i_1},{i_2},{i_3+1}}]}}$ (blue shades), ${\hat B_{[ {{i_1},{i_2+1},{i_3}}]}}$ (purple shades) and ${\hat B_{[ {{i_1+1},{i_2},{i_3}}]}}$ (red shades) for creating three-dimensional isoTNS of bond dimension $D=d$. The overlapping sites of these gates provide virtual bonds that connect tensor sites, leading to the desired cubic lattice structure of three-dimensional isoTNS.}
        \label{SGS_iso_hd}
\end{figure}

The RP-PEPS can also be defined in the same way, as starting from the source point, we group the unitaries into layers, and each layer of the gates acts on $q$-cubes across the boundary of the gate-acted region, which ballistically expand this region. The three-dimensional case is illustrated in \cref{SGS_iso_hd}(a). We also need to define the preferred direction as a permutation of directions ${1,2,...,q}$. If we assume the preferred directions is $\left( {1,2,...,q} \right)$, i.e. the first index of the position vector ${\vec v_\mu }$ grows first, we will get the following depth:
\begin{equation*}
{T_{{\rm{rp}}}} \approx \sum\limits_{i = 1}^q {L_p^{i - 1}{n_i}},
\end{equation*}
which lead to the scaling Eq.~(1) in the main text.

Compare to P-PEPS, it is less obvious to obtain the \textit{exact} circuit representation of isoTNS in higher dimension. Let us first draw some intuition of this representation from our two-dimension case. In the main text, the isoTNS with the geometry shown in Fig.~3(a7) in the main text have its bulk tensor equivalent to a three-qudit gate $\hat B_{[i,j]}$ acting on the sites $(i,j)$,
 $(i,j+1)$ and $(i+1,j+1)$. Notice that the geometry of $\hat B_{[i,j]}$ is not symmetric for the index $i,j$, which reflect the fact that $j$ is the preferred direction of the underlying RP-PEPS in Fig.~3(a).
                                                                   
It turns out that the tensor of $q$-dimensional isoTNS with bond dimension $D=d$ can be mapped to a unitary that acts on $q+1$ qudits, such that the number of the tensor indices $2q+1$ will match the number of incoming and outgoing indices of the unitary with one index acts on the reference state $\left| 0 \right\rangle $, which is also $2q+1$. For example, in the two-dimensional case [Fig.~3(b) in the main text], the PEPS tensor have 5 indices, corresponds to a 3-qudit unitary with a single input index as reference state $\left| 0 \right\rangle $.

 If we follow the same sequential ordering for the RP-PEPS discussed in this section, the unitary $\hat B_{[i_1,...,i_q]}$ that creates a single isoTNS tensor acts on the sites
\begin{equation*} \label{iso_hd}
\left\{ \begin{array}{l}
\left( {{i_1},...,{i_{q - 1}},{i_q}} \right),\left( {{i_1},...,{i_{q - 1}},{i_q} + 1} \right),\\
\left( {{i_1},...,{i_{q - 1}} + 1,{i_q} + 1} \right),...,\left( {{i_1} + 1,...,{i_{q - 1}} + 1,{i_q} + 1} \right)
\end{array} \right\}.
\end{equation*}

For example, to create three-dimensional isoTNS, we apply four-qudit unitary $\hat B_{[i_1,i_2,i_3]}$ that acts on 
\begin{equation} \label{}
	\left\{ \begin{array}{l}
\left( {{i_1},{i_2},{i_3}} \right),\left( {{i_1},{i_2},{i_3} + 1} \right),\\
\left( {{i_1},{i_2} + 1,{i_3} + 1} \right),\left( {{i_1} + 1,{i_2} + 1,{i_3} + 1} \right)
\end{array} \right\}.
\end{equation}

Several adjacent unitaries for preparing isoTNS of bond dimension $D=d$ are shown in \cref{SGS_iso_hd}(b) with different colors. The geometric overlapping of these unitaries gives rise to an isoTNS in a regular three-dimensional lattice, and it is clear that isoTNS $\subset$ RP-PEPS also holds in higher dimensions. \cref{SGS_iso_hd}(b) also shows that, given a unitary $\hat B_{[i_1,i_2,i_3]}$, its adjacent unitaries $\left\{ {{{\hat B}_{[{i_1} + 1,{i_2},{i_3}]}},{{\hat B}_{[{i_1},{i_2} + 1,{i_3}]}},{{\hat B}_{[{i_1},{i_2},{i_3} + 1]}}} \right\}$ commute with each other and can be implemented in the same layer of circuit. This eventually lead to a circuit depth for preparing $q$-dimensional isoTNS on a ${n_1} \times ... \times {n_q}$ lattice as [c.f.~Eq.~(8) in the main text]
\begin{equation*}
{T_{{\rm{iso}}}} \approx \sum\limits_{i = 1}^q {{n_i}},
\end{equation*}
which is slightly smaller than that for preparing RP-PEPS by a factor of $L^{q-1}_p$.

At last, one can also generalize the photon generation protocol to higher dimensions in the same way. To prepare a $q$-dimensional photonic lattice, we need a $(q-1)$-dimensional array where each site consists of an $D=d^{L_p-1}$-level ancilla and an $d$-level emitter as shown in Fig.~4 in the main text. Together the photon emission of emitters, one can operate the same sequential generation protocol as shown in Fig.~4 in the main text, where we replace the unitaries by $\hat U_{[i_1,...,i_q]}$ ($\hat B_{[i_1,...,i_q]}$) to prepare $q$-dimensional RP-PEPS (isoTNS).

\section{Proof of SGS $\subset$ isoTNS [c.f.~Eq.~(11) in the main text]}
\begin{figure}[h!]
	\centering
	\includegraphics[width=0.48\textwidth]{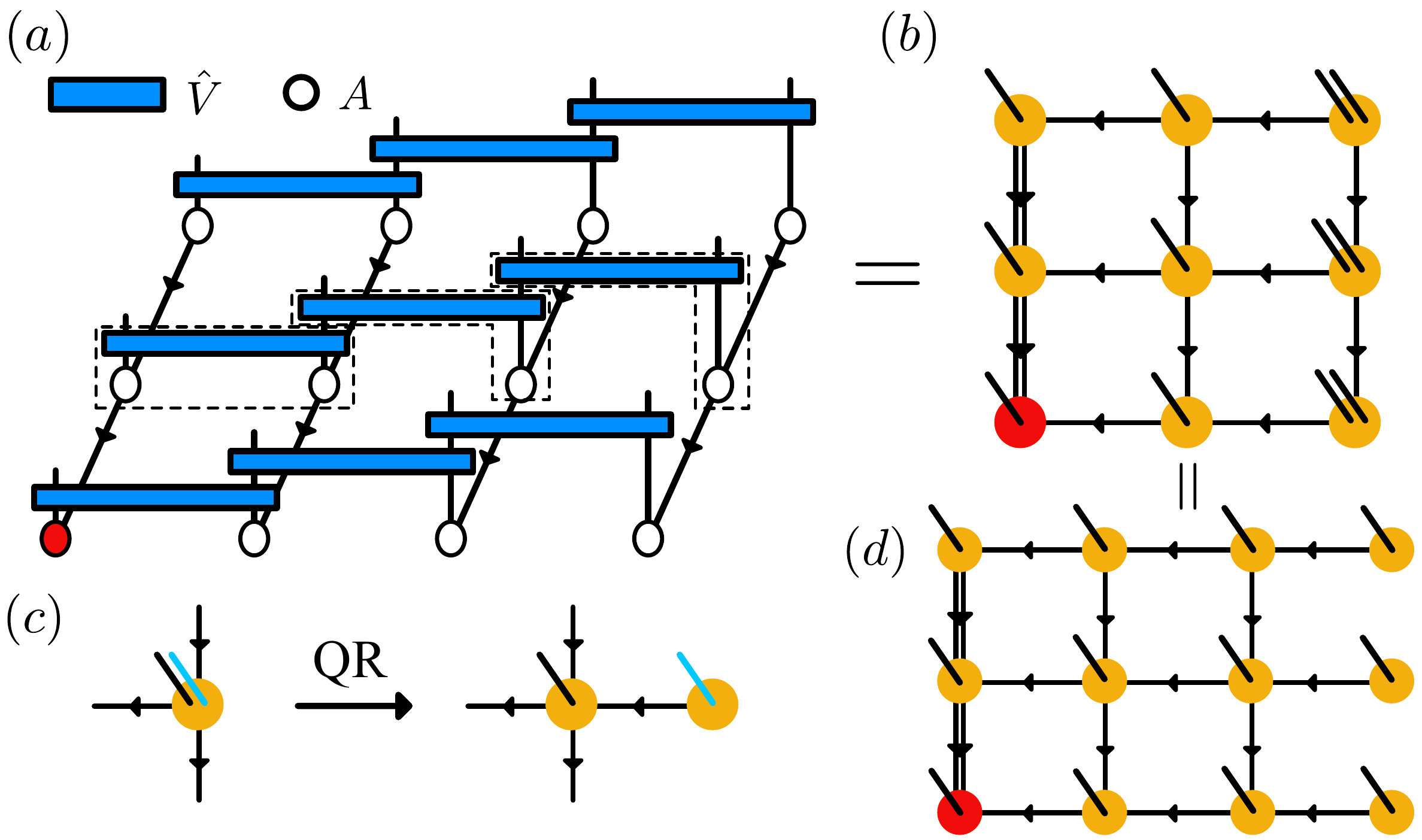}
        \caption{(a) The two-dimensional SGS (also see Fig.~2a in the main text). Here we further use arrows to denote the isometry conditions of MPS tensors in canonical form, to imply its relation with isoTNS. The OC of the corresponding isoTNS is shown as the red dot. (b) The SGS can be written as an isoTNS, with the OC marked by the red dot, and the structure of bonds shown in the figure. (c) One can do QR decomposition of tensors in panel (b) that have more physical indices, where we denote the index that will be decomposed out with blue color. This will result in multiple tensors, where each one satisfies an isometry condition. (d) After the QR decomposition, the SGS can be written as an isoTNS with uniform physical dimension $d$.}
        \label{SGS_peps}
\end{figure}

As already discussed in the main text, the SGS admit a PEPS representation~\cite{Banuls2008}. In this section, we prove SGS is a strict subclass of isoTNS using graphical notations. As introduced in the main text, the SGS is prepared by coupling unitaries $\{ {\hat V} \}$ on multiple lines of MPS. Without loss of generality, we can assume the MPS are all in the canonical form, with their orthogonality centers at the bottom of the lattice, as we show in \cref{SGS_peps}(a).

In SGS, the unitaries ${{\hat V}_{[ {i,j} ]}}$ in the bulk couple $L_p$ neighboring columns. The PEPS tensor inside the bulk of the lattice takes the form~\cite{Banuls2008}
\begin{equation} \label{SGS_tensor}
	\includegraphics[width=0.4\textwidth]{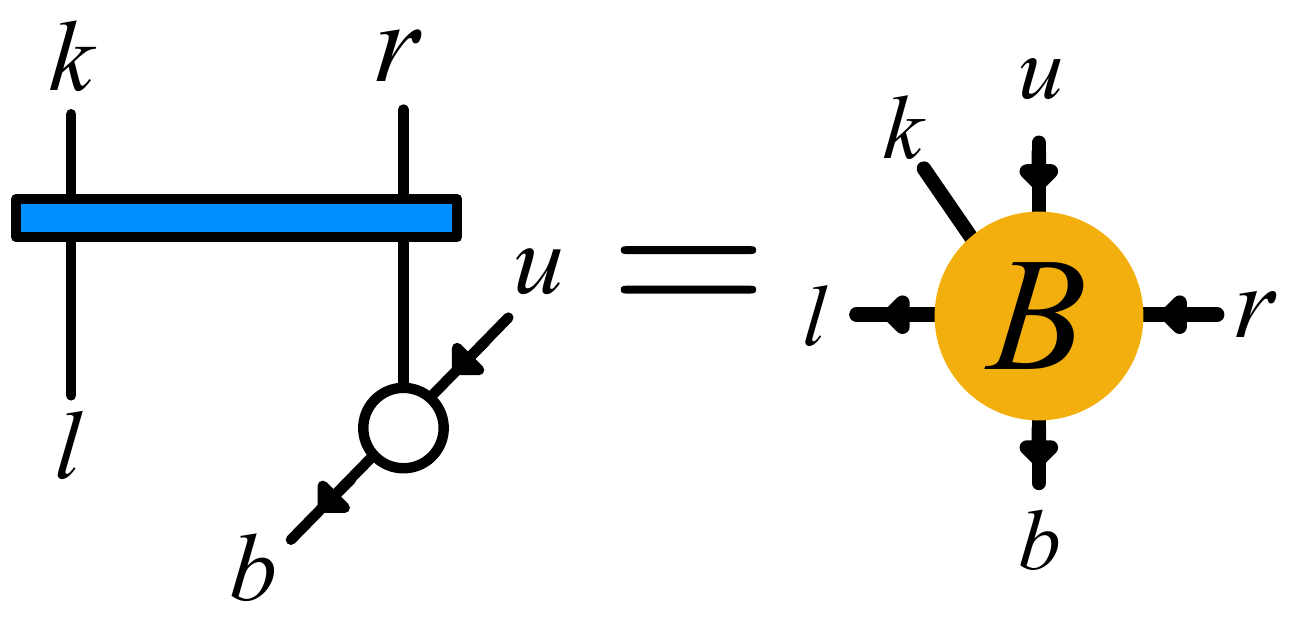}.
\end{equation}
Notice this relation is the same as that in Fig.~2(b) in the main text, and we further use the arrows to denote the isometry conditions. Depending on the bond dimension of the underlying MPS and the number of qub(d)its that the unitaries $\{\hat V\}$ acts on, the PEPS representation of SGS could have different bond dimensions in the horizontal and vertical bonds. Without loss of generality, we assume a uniform bond dimension $D=d^{L_p-1}$ in the bulk. Also notice that, at the left boundary of the two-dimensional lattice (the case of $L_p=2$ is shown in \cref{SGS_peps}a), the corresponding PEPS sites (denoted by the dashed box) have higher vertical bond dimension as $D^{L_p}$, and on the right boundary we group $L_p$ sites in each row (shown in the rightmost dashed box) as one, as PEPS sites with bond dimension $D$ and physical dimension $d^{L_p}$.

Thus SGS in \cref{SGS_peps}(a) can be mapped to PEPS of bond dimension $D'=D^{L_p} = d^{L_p(L_p-1)}$ and physical dimension $d'=d^{L_p}$, as shown in \cref{SGS_peps}(b) for the case of $L_p=2$. In \cref{SGS_peps}(b) we further use the isoTNS notations, as in the following we will show that the PEPS representation of arbitrary SGS is an isoTNS. We also point out that, one can do QR decomposition on the tensors that have multiple physical indices, to separate them to multiple tensor sites that have physical dimension $d$ and also satisfy isometry conditions (shown in \cref{SGS_peps}c). In this way, the isoTNS in \cref{SGS_peps}(b) can always be written as an isoTNS of a uniform bond dimension $d$ that have the same lattice size as the original SGS (\cref{SGS_peps}d).

\begin{proposition}
The PEPS representation of SGS [c.f.~\cref{SGS_peps}(b)] is an isoTNS.
\end{proposition}

Proof: One can directly check the isometry condition of the corresponding tensors in \cref{SGS_peps}(b). For the sites in the bulk, the unitarity of ${{\hat V}_{[ {i,j} ]}}$ and the canonical form of $A$ lead to
\begin{equation} \label{SGS_cond}
	\includegraphics[width=0.45\textwidth]{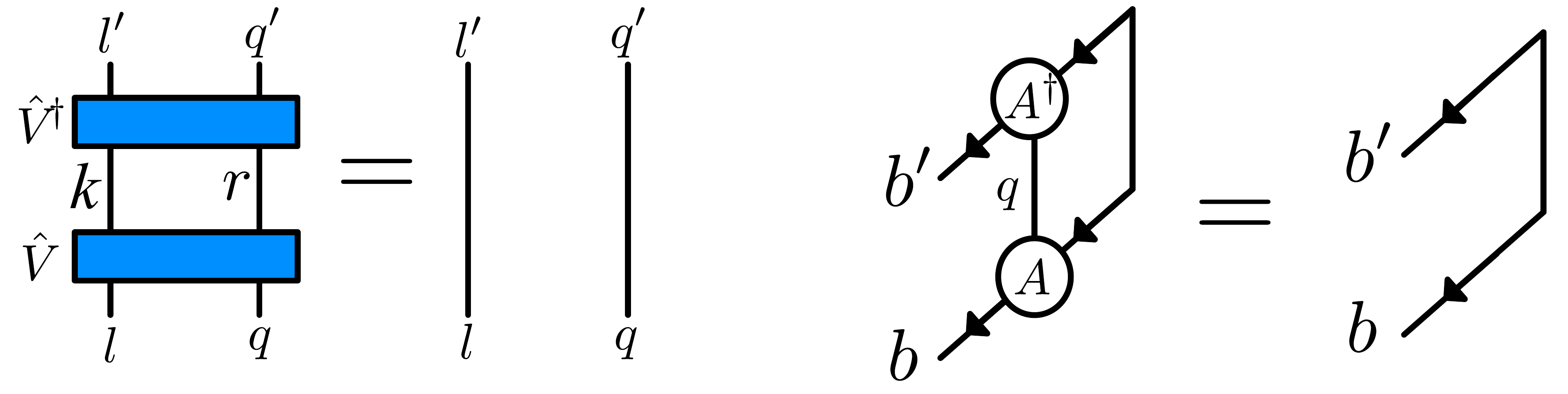},
\end{equation}
where the connected lines denotes the contraction of the corresponding indices. This directly leads to the isometry condition of SGS tensor  in the bulk [\cref{SGS_tensor}]:
\begin{equation}
	\includegraphics[width=0.48\textwidth]{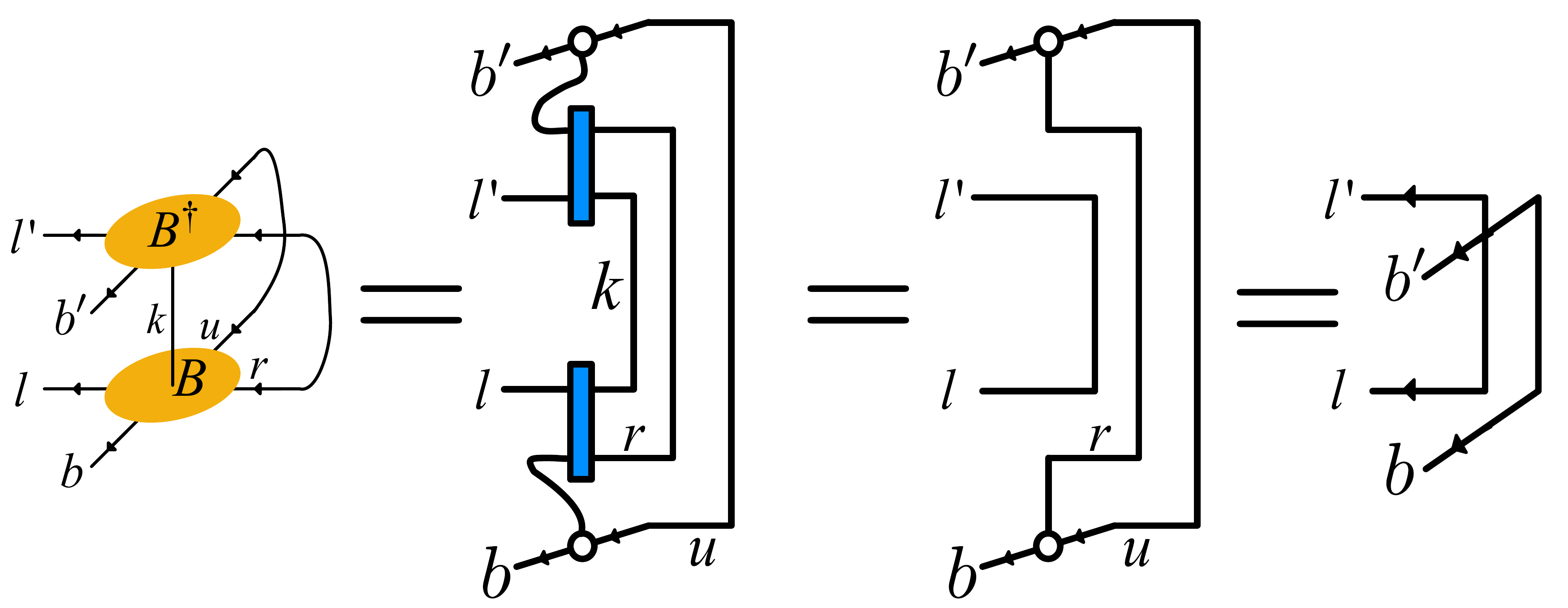}.
\end{equation}

Analogously, one can check all tensors in the boundaries also satisfy the corresponding isometry conditions, with arrow directions denoted in \cref{SGS_peps}(b). Thus the PEPS representation of SGS is itself an isoTNS (c.f.~\cref{SGS_peps}). $\square$

From the reverse direction, isoTNS is not necessarily a SGS. In the main text, we have qualitatively shown that the circuit representation of a tensor for SGS and isoTNS are quite similar, as they both acts on the same set of the qudits. In the case of bond dimension $D=d$, for SGS it contain two two-qudit unitaries [c.f.~Fig.~2(d) in the main text], while in the case of isoTNS the circuit is a generic three-qudit unitary [c.f.~Fig.~3(b) in the main text].

\begin{proposition}
A single PEPS tensor corresponding to a SGS site [c.f.~Eq.~(17)] does not cover arbitrary isoTNS tensor [c.f.~Fig.~3(b) in the main text] with the same isometry condition.
\end{proposition}
Proof: Let us consider the single PEPS tensor $B_{lurb}^k$ that have the same structure as that come from the SGS [c.f.~\cref{SGS_tensor}]. For simplicity, we consider the $L_p=2$ case, where $B_{lurb}^k$ has both physical dimension and bond dimension $d$, with the isometry condition shown by below \cref{iso_qr}. One can apply a QR decomposition to the tensor, getting
\begin{equation} \label{iso_qr}
	\includegraphics[width=0.4\textwidth]{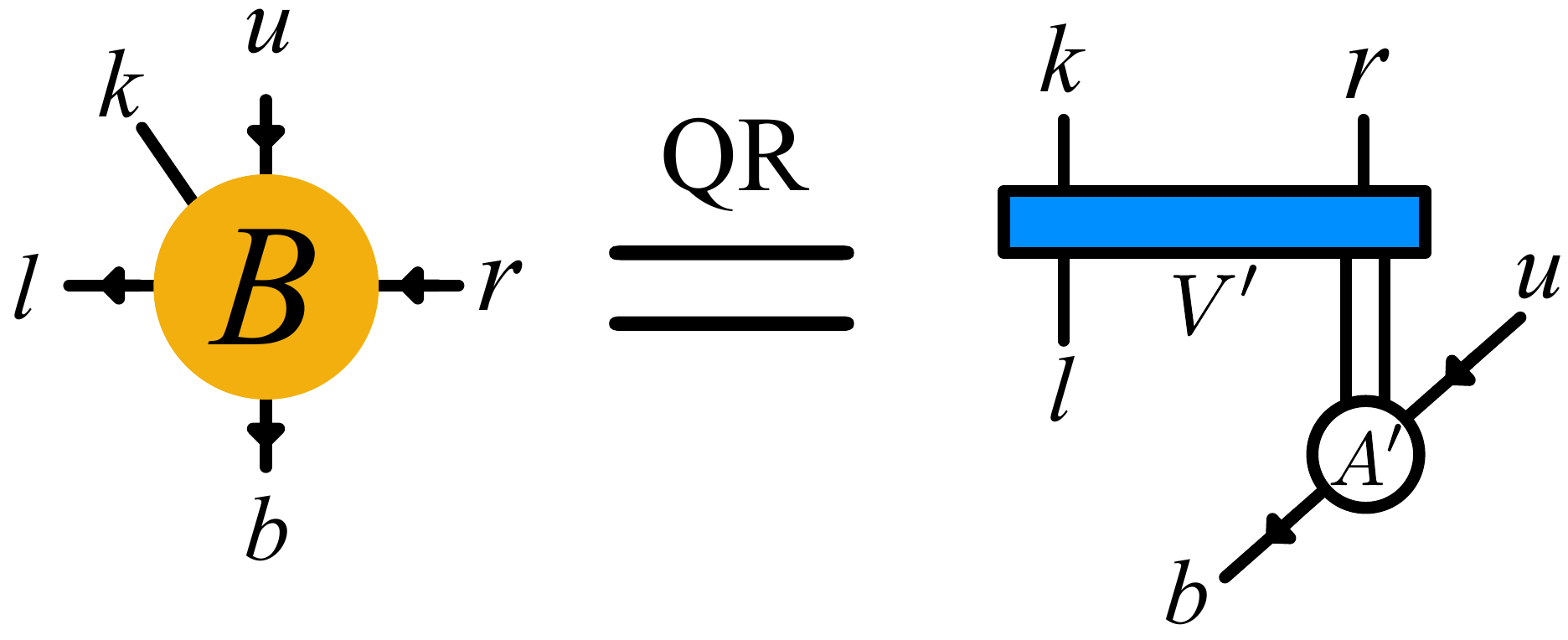},
\end{equation}
where the $V'$ part is an isometry (due to the QR decomposition), and each black line denote a bond of dimension $d$. Thus, an arbitrary isoTNS tensor satisfying isometry condition is equivalently represented by arbitrary $V'$ and $A'$ that satisfy
\begin{equation} \label{iso_cond}
	\includegraphics[width=0.45\textwidth]{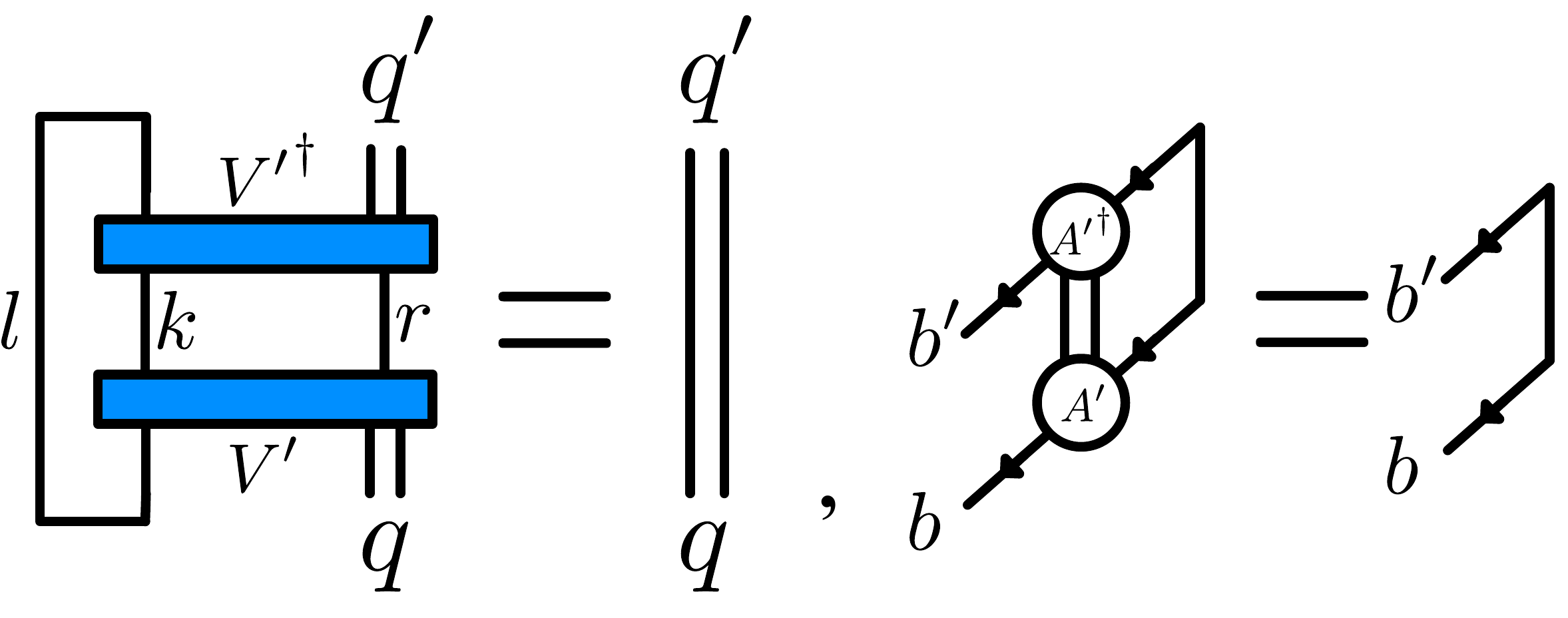}.
\end{equation}

It is clear that coming from an isoTNS, the resulting tensor coming from QR decomposition~[\cref{iso_cond}] does not necessarily satisfy the condition for SGS~\cref{SGS_cond}. They differ in the following points:
i) For the $V$ part of SGS, one requires a full identity on the indices $l$ and $q$ as shown in \cref{SGS_cond}. However, for the $V'$ part of isoTNS, we only require an identity relation on $q$ when we contract the other three indices $l,k,r$. ii) The index $q$ that connects two tensors $V'$ and $A'$ for isoTNS can generally have dimension $d^2$, while the one for SGS can only have the dimension $d$. The above point i) is particularly important, since even if one arbitrarily increases the bond dimension of the SGS, one can still find isoTNS that does not have an identity condition on the index $l$, thus makes SGS unable to cover the full class of isoTNS with any non-trivial bond dimension $D\ge 2$. $\square$

Thus SGS is a strict subclass of isoTNS. More precisely, we have [c.f.~Eq.~(11) in the main text]:
	\begin{equation*}
{\rm{SGS}}_{m,n}^{{L_p}} \subset {\rm{isoTNS}}_{m,n}^{{d^{L_p(L_p-1)}},d}.
\end{equation*}
Here we have used the PEPS representation of SGS [c.f.~\cref{SGS_peps}(d)]. We also point out that, here the bond dimension $D'={d^{L_p(L_p-1)}}$ of the isoTNS aims to capture the bond dimension at the boundary of the lattice, however, the majority of the sites in the bulk of the lattice just have $D={d^{L_p-1}}$, thus isoTNS of $D={d^{L_p-1}}$ `almost' cover the class of SGS of circuit length $L_p$.

The difference between SGS and isoTNS is reflected in their qualitative difference, that SGS admits efficient computation of local correlators at an arbitrary location of the lattice~\cite{Banuls2008}, however, isoTNS can only shift OC approximately~\cite{Zaletel2020}, thus compute local correlators at an arbitrary location only in an approximate way.

\section{Proof of F-PEPS $\subset$ P-PEPS [Eq.~(10) in the main text]}

\begin{figure}[h!]
	\centering
	\includegraphics[width=0.48\textwidth]{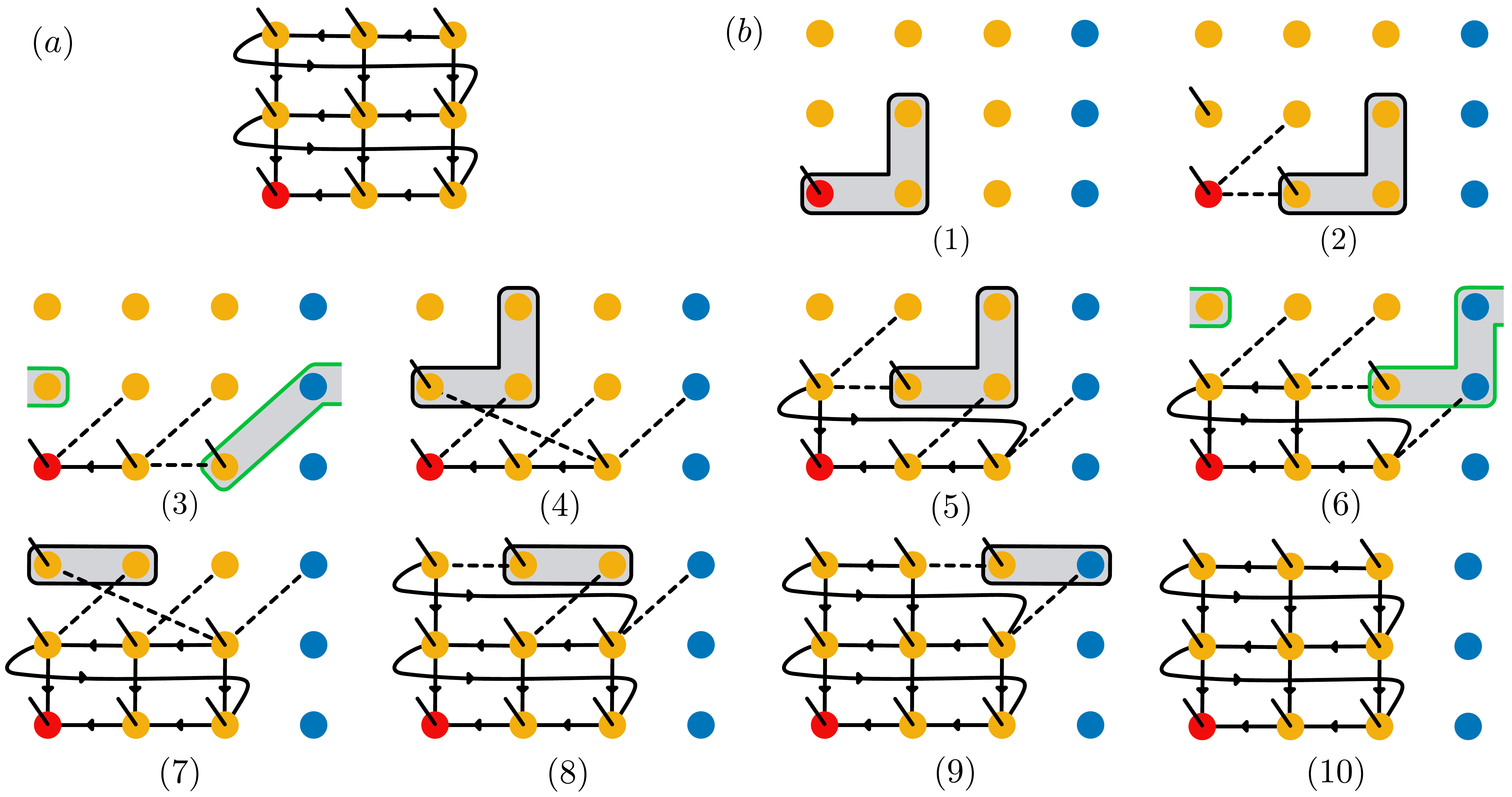}
        \caption{(a) The F-PEPS can be viewed as isoTNS with shifted periodic boundary condition. Here we use the same notation as that in Fig.~3(a) in the main text. (b) The sequential preparation of F-PEPS of bond dimension $D=d$ in an lattice. Here the blue dots denote the sites used as the ancilla.}
        \label{fp_circuit}
\end{figure}

The F-PEPS proposed in Ref.~\cite{Pichler2017} is a subclass of PEPS produced by photon feedback from a single emitter. It is clear from Ref.~\cite{Pichler2017} that, F-PEPS is a PEPS with shifted PBC, where each tensor satisfies an isometry condition. Thus, we can represent F-PEPS as isoTNS with a shifted PBC shown in \cref{fp_circuit}(a), where the isometry conditions are denoted by the arrows. In this section, we will show the preparation protocol of F-PEPS on lattices of stationary qub(d)its using a sequential circuit, which will naturally serve as proof of the fact that F-PEPS $\subset$ P-PEPS [Eq.~(10) in the main text].

The direction of the arrows in the isoTNS representation of F-PEPS
[c.f.~\cref{fp_circuit}(a)] indicates a full sequential ordering of its preparation circuit. In particular, the site on the boundary of each row has to be prepared earlier than the site on the boundary of the next row. Thus unlike the preparation of isoTNS with OBC [c.f.~Fig.~3(a) in the main text], the preparation of F-PEPS does not allow the gates to be applied simultaneously along different directions.

We can prepare F-PEPS with a full sequential circuit shown in \cref{fp_circuit}(b). Compare to the isoTNS preparation protocol [c.f.~Fig.~3(a) in the main text], here we modify the shape of gates that create sites on the right boundary of each row (denoted with green boxes in steps 3 and 6), such that these gates now also act on the first site of the next row. The coupling of different boundaries creates additional virtual bonds of the resulting isoTNS, thus give us an arbitrary isoTNS of bond dimension $D=d$ of the geometry shown in \cref{fp_circuit}(b10), with shifted PBC. Here we have used the sites on the right boundary of the lattice as ancillas. 

Similar to the isoTNS case [c.f.~\cref{iso_anci}], to create arbitrary F-PEPS with larger bond dimension $D$, we need to let the neighboring gates to have $s = \left\lceil {{{\log }_d}D} \right\rceil$ common sites, thus require $s$ columns and $s-1$ rows of ancillas (later we will put $s$ rows of ancillas, to ensure the gates are covered by plaquette unitaries). Also, the sequential circuit of the maximal size are those that couples to boundaries of the lattice( shown as green boxes in \cref{fp_circuit}), and they can be contained in plaquette of side length $L_p = 2s+1$. With that, it is clear that [c.f.~Eq.~(10) in the main text]
\begin{equation*}
{\textrm{F-PEPS}}_{n \times m}^{D,d} \subset {\textrm{P-PEPS}}_{(n + s) \times (m+s)}^{2s + 1}.
\end{equation*}

The above construction shows that F-PEPS $\subset$ P-PEPS, and serve as the way to create them. To create a F-PEPS on $N=n\times m$ lattice, 
the circuit depth is $T=O(N)$. We also point out that isoTNS with other kinds of periodic boundary conditions can be prepared in the same way.

\section{Preparing isoTNS and F-PEPS of flying qub(d)its}

As shown in the main text, each isoTNS tensor can be mapped to a `L'-shape unitary $\hat B_{[i,j]}$ [c.f.~Eq.~(7) in the main text]. Thus using the ancilla-emitter array setup shown in Fig.~4 in the main text, we are able to create photonic isoTNS.

 \begin{figure}[tb]
	\centering
	\includegraphics[width=0.48\textwidth]{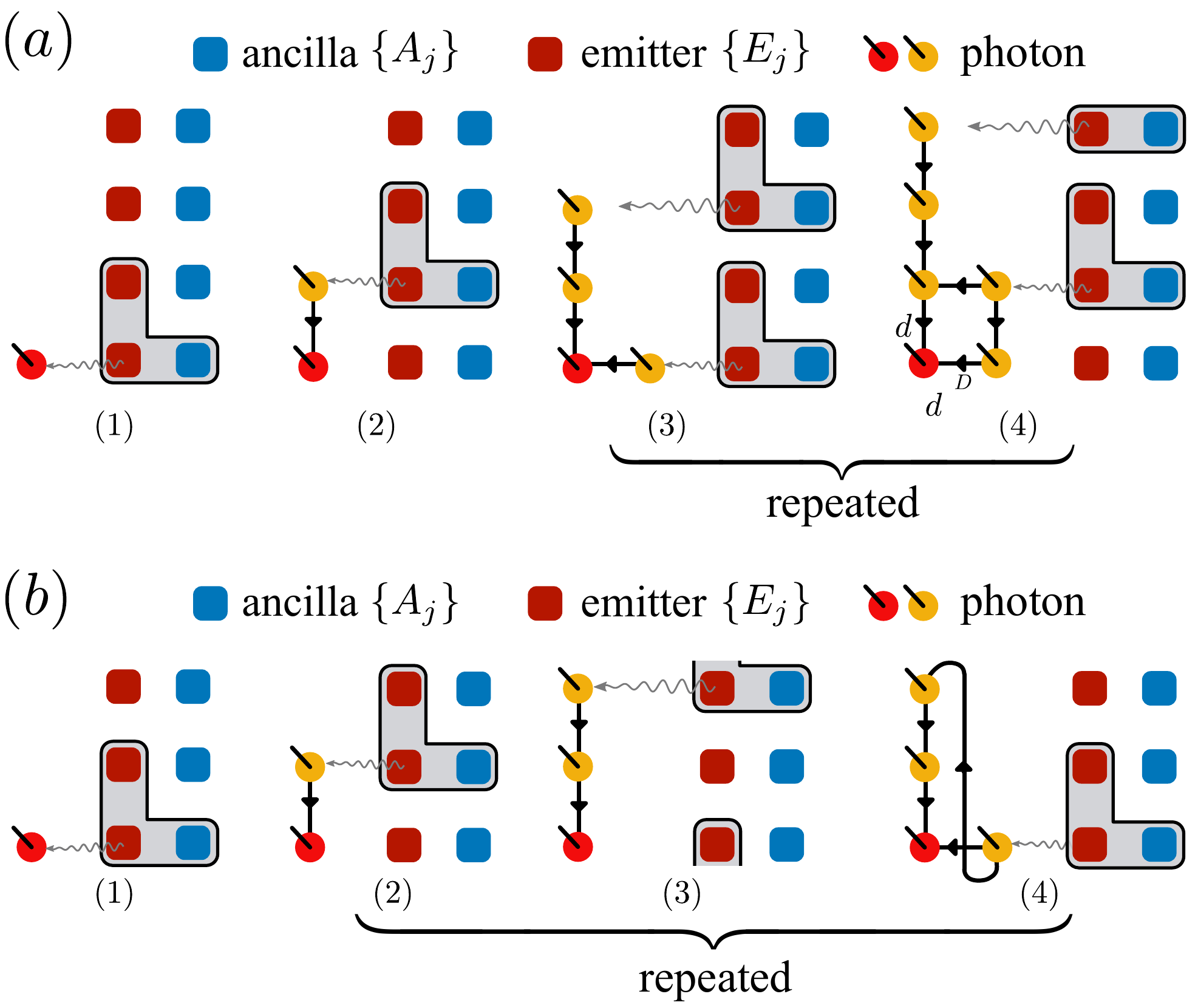}
        \caption{Generation of photonic isoTNS using an array of coupled sequential photon sources. (a) To prepare isoTNS with open boundary condition, one apply the same procedure of the RP-PEPS generation shown in Fig.~4 in the main text, with the `L'-shape gates. (b) By coupling the two ends of the array of sequential photon sources with `L'-shape gate (shown in the step 3), one can prepare F-PEPS, which are photonic isoTNS with shifted periodic boundary condition.}
                \label{isop_si_gen}
\end{figure}

The procedure to produce photonic isoTNS of bond dimension $D=d$ with the open boundary condition is shown in \cref{isop_si_gen}(a). Here we follow the RP-PEPS generation protocol in Fig.~4 in the main text, and the unitary $\hat B_{[i,j]}$ that prepares the $[i,j]$-th isoTNS site acts on the ancilla $A_j$ and emitters $E_j, E_{j+1}$. Due to the sequential nature of the protocol, the OC sits on the first photonic qudit being produced. One can also increase the bond dimension of the resulting isoTNS in the same way as shown in \cref{bond_incr}(a), by increasing the ancilla dimension and increase the length of the `L'-shape gate along the emitters. Also, similar to the matter-based isoTNS case [c.f.~\cref{iso_anci}(b)], we need to put additional $s-1$ qudits (each of the dimension $d$) with $s = \left\lceil {{{\log }_d}D} \right\rceil $ on the top or bottom boundary of the array, which will be used to create virtual bonds of the isoTNS tensors.

In this case, this photonic isoTNS is again contained in the photonic RP-PEPS on a slightly larger lattice.

At last, by another simple modification of the above protocol, our platform can also prepare photonic F-PEPS, which are isoTNS with \textit{shifted} periodic boundary conditions (sPBC). In \cref{isop_si_gen}(b), we show the steps to prepare F-PEPS. The key point is step (3), which applies a gate that couples two boundaries of the lattice, providing the virtual bond for sPBC shown in step (4). This indicates that the class of states that our ancilla-emitter array setup can prepare is strictly larger than that the non-Markovian feedback approaches can prepare~\cite{Pichler2017, Dhand2018,Xu2018,Wan2020,Zhan2020,Shi2021,Bombin2021}.

Compared to the non-Markovian feedback approaches~\cite{Pichler2017, Dhand2018,Xu2018,Wan2020,Zhan2020,Shi2021,Bombin2021}, our array-based approach does not require interaction with the emitted photons, therefore is easier to implement experimentally. Specifically, it allows one to implement generic unitaries and makes it possible to increase the complexity of the state by acting on more qub(d)its.

\bibliography{library.bib}
\end{document}